\documentclass[]{interact}

\usepackage[utf8]{inputenc}
\usepackage[T1]{fontenc}
\usepackage[english]{babel}
\usepackage{textcomp}
\usepackage{latexsym, amsmath, amssymb}
\usepackage{booktabs, multirow}
\usepackage[version=4]{mhchem}

\usepackage[numbers,sort&compress]{natbib}
\bibpunct[, ]{[}{]}{,}{n}{,}{,}

\usepackage{epstopdf}
\usepackage{graphicx}
\usepackage{color}
\usepackage[caption=false]{subfig}

\usepackage{mathtools}
\usepackage{exscale, relsize}
\usepackage[retainorgcmds]{IEEEtrantools}

\usepackage{siunitx}
\DeclareSIUnit\atm{atm}

\usepackage[breaklinks=true, linkcolor=blue, citecolor=blue, colorlinks=true]{hyperref}

\usepackage{xargs}
\usepackage[pdftex,usenames,dvipsnames]{xcolor}

\makeatletter
\newcounter{reaction}
\renewcommand\thereaction{R\arabic{reaction}}
\newcommand\reactiontag%
  {\refstepcounter{reaction}\tag{\thereaction}}

\makeatother

\begin{document}

\articletype{ARTICLE}

\title{Smoldering combustion in cellulose and hemicellulose mixtures: 
Examining the roles of density, fuel composition, oxygen concentration, 
and moisture content}


\author{
\name{W.~Jayani Jayasuriya, Tejas Chandrashekhar Mulky, and 
Kyle E.~Niemeyer\thanks{CONTACT K.~E.~Niemeyer. Email: kyle.niemeyer@oregonstate.edu}}
\affil{School of Mechanical, Industrial, and Manufacturing Engineering, 
Oregon State University, Corvallis, OR, USA}
}

\maketitle

\begin{abstract}{
  Smoldering combustion plays a key role in wildfires in forests, grasslands,
  and peatlands due to its common occurrence in porous fuels like peat and duff.
  As a consequence, understanding smoldering behavior in these fuels is crucial.
  Such fuels are generally composed of cellulose, hemicellulose, and lignin.
  Here we present an updated computational model for simulating smoldering combustion in
  cellulose and hemicellulose mixtures. We used this model to examine changes in smoldering
  propagation speed and peak temperatures with varying fuel composition and density. For a given fuel
  composition, increases in density decrease the propagation speed and increase mean peak temperature;
  for a given density, increases in hemicellulose content increase both propagation speed and
  peak temperature. 
  We also examined the role of natural fuel expansion with the addition of water. 
  Without expansion, addition of moisture
  content reduces the propagation speed primarily due to increasing (wet) fuel density.
  However, with fuel expansion similar to that observed in peat, the propagation speed increases
  due to the overall drop in fuel density. Finally, we studied the influence of fuel composition
  on critical moisture content of ignition and extinction: mixtures dominated
  by hemicellulose have 10\% higher critical moisture content due to
  the increase in peak temperature.
}
\end{abstract}



\section{Introduction}
\label{p2:Introduction}

Wildland fires lead to human, environmental, and ecological hazards.
Global climate change has and will continue to cause increases in the occurrence 
of droughts, which will in turn lead to an increasing frequency of wildland 
fires~\citep{LIU2010685,Watts2013}. Combustion in wildland fires, in general, is 
dominated by either flaming or smoldering combustion.
Both types of combustion have different characteristics and can be hazardous in their own way, 
but flaming combustion has historically received more research compared with smoldering.
However, as Rein~\cite{Rein2009} discussed, smoldering combustion has recently become more recognized 
as a major fire hazard, resulting in increasing interest in understanding this phenomenon.

Compared with flaming combustion, smoldering can persist longer and under conditions that 
would extinguish flames. This characteristic of smoldering combustion allows it to penetrate 
deeper into the soil compared with flaming combustion, which generally causes shallower 
burns~\citep{Hartford,Rein-ch2}. Thus, smoldering can actually cause greater destruction 
in affected ecosystems. 
Smoldering also emits a large number of pollutants such as carbon monoxide (\ce{CO}), 
volatile organic compounds, polycyclic aromatic hydrocarbons, and particulate matter, 
since it operates at lower temperatures than flaming combustion. 
Smoldering occurs most commonly in 
porous fuels like peat, woody fuels, muck, and forest duff~\citep{Watts2013}. Such fuels are abundant in forests, 
making it important to understand smoldering combustion in these types of fuels.
Woody fuels and biomass generally consist of cellulose, hemicellulose, and lignin in varying proportions,
which pyrolyze at different temperatures as shown by Ranzi et al.~\citep{ranzi_chemical_2008,ranzi_kinetic_2014}.
Yang et al.~\cite{YANG20071781} found that, among the three, 
hemicellulose pyrolyzes earliest, at temperatures of \SIrange{220}{315}{\degreeCelsius}, 
cellulose undergoes pyrolysis at temperatures of \SIrange{315}{400}{\degreeCelsius},
and finally lignin pyrolyzes at temperatures of \SIrange{150}{900}{\degreeCelsius}.
Anca-Couce et al.~\cite{anca-couce_smouldering_2012} showed similar trends in pyrolysis 
of these three constituents in their thermogravimetric analysis of pine wood.
In addition, these fuel constituents produce different amounts of 
char~\citep{Huang2016,Kashiwagi1992,CAGNON2009292}.
Smoldering combustion is generally modeled using a set of global reactions,
which include fuel pyrolysis and char oxidation~\citep{Rein2016_book,anca-couce_smouldering_2012}.
Differences in fuel composition thus may lead to significant differences in smoldering characteristics.
This motivates our detailed study looking into how varying fuel composition affects smoldering 
characteristics.

Along with fuel composition, the other parameters that could affect smoldering 
propagation are density and moisture content. Huang and Rein~\cite{Huang2017} found 
that increasing the density of peat by 40\% reduces the downward 
propagation speed by approximately 40\%.
However, no (computational) studies have looked into how changes in density affect 
smoldering speed and temperatures in fuel mixtures of cellulose and hemicellulose.
In contrast, regarding the effects of moisture content, Huang and Rein~\cite{Huang2017} 
studied how moisture content affects the propagation speed of peat and observed an increase
in downward propagation speed with moisture content, due to expansion of the peat.
Recently, Smucker et al.~\cite{Smucker:2019sm,Smucker2021correction} experimentally observed that smoldering propagation
speed in mixtures of cellulose and hemicellulose decreases with density, and attributed 
this to oxygen availability.
They also found that propagation speed increases with additional hemicellulose content in fuel,
attributed to faster pyrolysis with addition of hemicellulose, from its lower activation energy 
and higher heat release.

Critical moisture content is the highest moisture content above which smoldering combustion 
cannot self-sustain. Garlough and Keyes~\cite{Garlough} experimentally studied ponderosa 
pine duff and found that fuel consumption decreases after reaching critical moisture content 
of 57 and 102\% on the upper and lower duff, respectively.
Frandsen showed experimentally that duff's critical moisture content of ignition drops with 
inorganic content~\cite{Frandsen1987,Frandsen1997}.
Huang and Rein~\cite{huang_critical,Huang2015} found that natural peat's critical moisture 
contents of ignition and extinction are around 117\% and 250\%, respectively, but vary 
significantly depending upon the thickness of wet layer, dry layer, inorganic content, 
physical properties, and boundary conditions.
However, no studies have looked into the influence of the fuel composition on these threshold values.

In our prior work, we found that propagation speed increases as density drops or hemicellulose content
increases for mixtures of cellulose and hemicellulose~\citep{Tejas}. 
Based on prior theories in the literature, we hypothesized that oxygen availability 
causes the sensitivity to density,
and that adding hemicellulose increases propagation speed since it pyrolyzes faster.
However, that study did not include an in-depth analysis to examine the proposed hypotheses or their fundamental causes.
In addition, for validating the model with experimental results, we relied on a fixed temperature 
boundary condition, which overconstrained the model. 
Furthermore, our previous treatment of bulk density for validation case may not represent actual 
experimental conditions: we fixed the bulk density of hemicellulose and changed the bulk 
density of cellulose to match the mixture bulk density; 
in experiments, they change together~\citep{Blunck-comm,Smucker:2019sm,Smucker2021correction}.
The model used in that work did not predict ignition for bulk densities of less than
\SI{200}{\kilo\gram\per\meter^3} for 100\% cellulose, which disagrees with experimental 
observations~\citep{Smucker:2019sm,Smucker2021correction}.
Here, we use a more-appropriate boundary condition at the upper surface, 
allow the bulk density of the fuel components to vary independently, and
updated physical property values (e.g., particle surface area).
This study also expands on the analysis of the reasons behind observed trends in 
propagation speed and peak temperature, confirms the relationship between oxygen availability 
and density posited for peat by Huang and Rein~\cite{Huang2017}, 
confirms---and extend to general fuels---the observation by Huang and Rein~\cite{Huang2017} that 
moisture content increases downward smoldering in peat, and also examines the impact of 
fuel composition on critical moisture content of ignition and extinction.

Building on our prior work, this article presents an updated one-dimensional, 
transient computational model to simulate smoldering combustion in cellulose and hemicellulose 
mixtures.
First, we validate the model against a different experimental configuration that 
more closely matches the simulation, and use a heat-flux boundary condition. 
Following this model validation, we examine the effects of
varying density and fuel composition on smoldering propagation speed and peak temperature, and
perform an in-depth analysis to explain the observed trends. Next, we investigate the effects of
varying moisture content on smoldering propagation speed and temperature, including and excluding the
contribution of fuel expansion with the addition of water. Finally, we identify how varying fuel composition
affects the critical moisture content of ignition and extinction.

\section{Computational model}
In this article, we study downward propagation of smoldering using a one-dimensional transient
model following approaches of past studies~\citep{Tejas}. This model was developed using Gpyro~\citep{gpyro:0.7}.
We performed simulations with a spatial cell size $(\Delta z)$ of \SI{1e-4}{\meter} and an initial
time step of \SI{0.05}{\second}. We based this selection of cell size on our previous work,
where we showed that further increasing resolution has little impact on global quantities of interest~\citep{Tejas}.

\subsection{Governing equations}
To model smoldering combustion, we use Gpyro v0.700~\citep{Lautenberger2009,gpyro:0.7} 
to solve the transient governing equations:
condensed-phase mass conservation~\eqref{condensed-phase mass_1},
condensed-phase species conservation~\eqref{condensed-phase species_1},
gas-phase mass conservation~\eqref{gas-phase mass_1},
gas-phase species conservation~\eqref{gas-phase species_1},
condensed-phase energy conservation~\eqref{condensed-phase energy_1},
gas-phase momentum conservation~\eqref{gas-phase momentum_1}, and
gas-phase energy conservation~\eqref{gas-phase energy_1}; 
the ideal gas equation of state~\eqref{ideal-gas law_1}
is needed to close the set of equations.
Lautenberger and Fernandez-Pello~\cite{Lautenberger2009} provide more details about Gpyro.
For completeness, the governing equations are:
{\allowdisplaybreaks \begin{IEEEeqnarray}{rCl}
\frac{\partial\overline{\rho}}{\partial t} &=& - \dot{\omega}^{\prime\prime\prime}_{fg} \;, \label{condensed-phase mass_1} \\
\frac{\partial(\overline{\rho}Y_i)}{\partial t} &=&  \dot{\omega}^{\prime\prime\prime}_{fi} - \dot{\omega}^{\prime\prime\prime}_{di} \;, \label{condensed-phase species_1} \\
\frac{\partial(\rho_g\overline{\psi})}{\partial t} + \frac{\partial \dot{m}^{\prime\prime}}{\partial{z}} &=& \dot{\omega}^{\prime\prime\prime}_{fg} \;, \label{gas-phase mass_1} \\
\frac{\partial (\rho_g\overline{\psi}Y_j)}{\partial t} + \frac{\partial{(\dot{m}^{\prime\prime} Y_j)}}{\partial z} &=& -\frac{\partial}{\partial z}(\overline{\psi}\rho_gD\frac{\partial Y_j}{\partial z})+ \dot{\omega}^{\prime\prime\prime}_{fj} - \dot{\omega}^{\prime\prime\prime}_{dj} \;, \label{gas-phase species_1} \\
\frac{\partial(\overline{\rho}\overline{h})}{\partial t} &=&
\frac{\partial}{\partial z}(\overline{k}\frac{\partial T}{\partial z})- \dot{Q}^{\prime\prime\prime}_{s-g}+ \sum_{k=1}^{K} \dot{Q}^{\prime\prime\prime}_{s,k}- \frac{\partial \dot q^{''}_r}{\partial z} \nonumber \\
&& +\: \sum_{i=1}^{M} ((\dot \omega ^{'''}_{fi}-\dot{\omega}^{\prime\prime\prime}_{di}) h_i) \;, \label{condensed-phase energy_1} \\
\dot{m}^{\prime\prime} &=& -\frac{\overline{K}}{v}\frac{\partial P}{\partial z} \;,\text{ and} \label{gas-phase momentum_1} \\
\frac{\partial(\overline{\psi}\rho_g\overline{h_g})}{\partial t}+\frac{\partial(\dot m^{''}_{z} \overline{h_g})}{\partial z} &=&
\frac{\partial}{\partial z}(\overline{\psi}\rho_{g}D\frac{\partial\overline{h_g}}{\partial z})+ h_{cv}(T-T_{g}) \nonumber \\
&& +\: \sum_{j=1}^{N} (\dot \omega ^{'''}_{s,fj}-\dot{\omega}^{\prime\prime\prime}_{s,dj}) h^*_{g,j}+\dot{Q}^{\prime\prime\prime}_{s-g} \;, \label{gas-phase energy_1} \\
P \overline{M} &=& \rho_g R T_g \;, \label{ideal-gas law_1}
\end{IEEEeqnarray}}%
where $\rho$ is the density, $M$ is the number of condensed-phase species; $X$ is the volume fraction;
$\dot{\omega}^{\prime\prime\prime}$ is the reaction rate; $T$ is the temperature;
$Y_j$ is the $j$th species mass fraction; $\psi$ is the porosity;
$K$ is the permeability/number of reactions; $h_{cv}$ is the volumetric heat transfer coefficient;
$\overline{M}$ is the mean molecular mass obtained from local volume fractions of all gaseous species;
$\dot{q}^{\prime\prime}_r$ is the radiative heat-flux;
$\dot{Q}^{\prime\prime\prime}$ is the volumetric rate of heat release/absorption;
$R$ is the universal gas constant; $D$ is the diffusion coefficient; $h$ is the enthalpy;
$P$ is the pressure; subscripts $f$, $d$, $i$, $j$, $k$, $s$, and $g$ are formation, destruction,
condensed-phase species index, gas-phase species index, reaction index, solid, and gas; and
$^*$ indicates that gas-phase species enthalpy is calculated at condensed phase temperature.
The overbars over $\rho$, $\psi$, $K$, and $k$ mean an averaged value weighted by condensed-phase volume fraction,
while the overbar over $h$ indicates an averaged value weighted by condensed-phase mass fraction.

\subsection{Boundary conditions}
\label{Boundary conditions}

\begin{figure}[htbp]
\centering
\includegraphics[width= 0.4\linewidth]{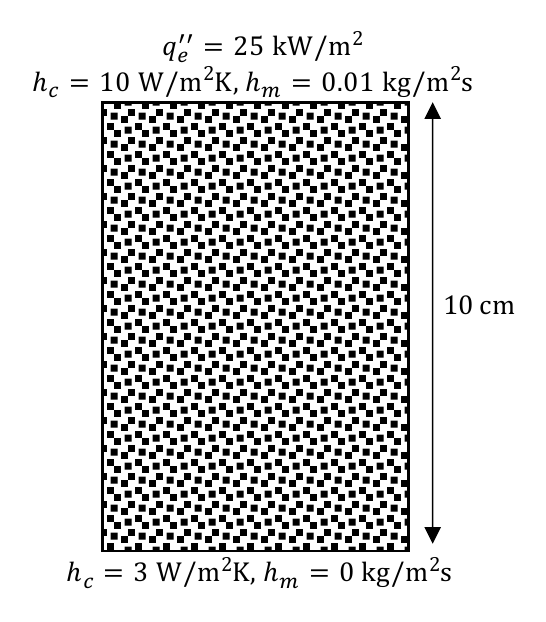}
\caption{Schematic illustration of the one-dimensional computational domain.}
\label{fig:domain}
\end{figure}

The top surface ($z=0$) of the domain was modeled as open to atmosphere while the bottom surface
($z=L$) was modeled as insulated to match the experimental setup, as Figure~\ref{fig:domain} shows.
The pressure ($P$) at the top surface was \SI{1}{\atm} and the ambient temperature was \SI{300}{\kelvin}.
On the top surface we set a convective heat transfer coefficient $(h_{c,0})$
as \SI{10}{\watt\per\meter^2\kelvin} using an empirical correlation of
$h_{c,z=0}=1.52\times T^{1/3}$ where $T = \SI{300}{\kelvin}$~\citep{Huang2015}.
At the upper surface we also set the mass-transfer coefficient $(h_{m,0})$ at
$\SI{0.02}{\kilogram\per\meter^2\sec}$ based on previous work~\citep{Huang2015}.
To ignite the sample we provided a heat flux $(\dot{q_{e}^{\prime\prime}})$ of \SI{25}{\kilo\watt\per\meter^2}
for \SI{20}{\minute} at the top boundary to establish self-sustained smoldering,
after which we removed the heat flux and established a convective--radiative balance
at the top surface (e.g., for $t >\SI{20}{\minute}$):
%
\begin{align}
\left. -\overline{k}\frac{\partial T}{\partial z} \right\rvert_{z=0} &= -h_{c0}(T_{z=0}-T_{\infty})+\overline{\epsilon}\dot{q}_{e}^{\prime\prime} - \overline{\epsilon}\sigma(T_{z=0}^4-T_{\infty} ^{4})] \;, \label{bc_top_surface_2_1} \\
\left. -\overline{k}\frac{\partial T}{\partial z} \right\rvert_{z=0} &= -h_{c0}(T_{z=0}-T_{\infty})-\overline{\epsilon}\sigma(T^4_{z=0}-T_{\infty} ^{4}) \;, \label{bc_top_surface_3_1} \\
-\left. \left( \overline{\psi}\rho_g D \frac{\partial Y_j}{\partial z} \right) \right\rvert_{z=0} &= h_{m0} \left( Y_{j\infty}-Y_{j} \rvert_{z=0} \right) \label{bc_top_surface_4_1} \;, \text{ and} \\
P\rvert_{z=0} &= P_{\infty} \;. \label{bc_top_surface_5_1}
\end{align}
We applied these boundary conditions for all simulations, except those looking at the effects of
varying moisture content on propagation speed (Sec.~\ref{sec:moisturecontent})
where we set a constant heat flux throughout the simulation to guarantee ignition
at higher moisture contents.

For the bottom surface we set a heat-transfer coefficient ($h_{c,L}$) of
$\SI{3}{\watt\per\meter^2\kelvin}$ to account for losses through the insulation.
The mass flux $(\dot{m}^{\prime\prime})$ was set to zero at the bottom surface.
The equations used for boundary conditions on the bottom surface are
\begin{align}
\left. -\overline{k}\frac{\partial T}{\partial z} \right\rvert_{z=L} &= -h_{cL}(T\rvert_{z=L}-T_{\infty}) \label{bc_back_surface_1_1} \;, \text{ and}\\
\left. -\left(\overline{\psi} \rho_g D \frac{\partial Y_j}{\partial z}\right) \right\rvert_{z=L} &= 0 \;. \label{bc_back_surface_2_1}
\end{align}

\subsection{Chemical kinetics}
\label{Chemical Kinetics}

Gpyro represents heterogeneous reactions as~\citep{Lautenberger2009}:
\begin{align}
    \ce{A_{k} + $\sum_{j=1}^{N} v^{\prime}_{j,k}$ gas j  &-> $v_{B,k}$ B_{k} + $\sum_{j=1}^{N} v^{\prime\prime}_{j,k}$ gas j} \;,
\end{align}
where $k$ represents the reaction number, \ce{A_k} and \ce{B_k} are condensed-phase species,
$v^{\prime}_{j,k}$ and $v^{\prime\prime}_{j,k}$ are the reactant and product stoichiometric
coefficients for \ce{gas j} in reaction $k$,
$v_{B,k}$ is the stoichiometric coefficient for condensed-phase species \ce{B} in reaction $k$,
and $N$ is the total number of gas-phase species.
The reaction rates are expressed in Arrhenius form:
\begin{equation}
  \dot\omega^{\prime\prime\prime}_{dA_k} =
  Z_k \frac{\left(\overline{\rho} Y_{A_k} \Delta z \right)_{\sum}}{\Delta z}
  \left(\frac{\overline{\rho} Y_{A_k} \Delta z}{\left( \overline{\rho} Y_{A_k} \Delta z \right)_{\sum}} \right)^{n_k}
  \times \exp{\left(-\frac{E_k}{RT}\right)} g(Y_{\ce{O2}}) \;,
\label{reaction_rate_11}
\end{equation}
where
\begin{equation}
\left( \overline{\rho} Y_{A_k} \Delta z \right)_{\sum} = \overline{\rho}Y_{A_k}\Delta z\rvert_{t=0} + \int_{0}^{t}\dot\omega{^{\prime\prime\prime}_{fi}}(\tau)\Delta z(\tau) d\tau \;,
\end{equation}
$Z$ is the pre-exponential factor, $E$ is the activation energy, $n$ is the order of reaction,
subscript $dA$ stands for destruction of species \ce{A}, and subscripts $k$, $f$, and $i$ are 
reaction index, formation, and condensed-phase species index.
In Eq.~\eqref{reaction_rate_11}, for inert atmosphere $g(Y_{O_2}) =1$
and when oxygen is available $g(Y_{O_2}) =(1 + Y_{O_2})^{n_{O_2,k}}-1$~\citep{Lautenberger2009}.

We represent the smoldering process with a system of global pyrolysis and oxidation
reactions~\citep{Anca-Couce2012,Rein2016_book}, using the model developed by
Huang and Rein~\cite{Huang2014,Huang2016} for smoldering of the mixtures of interest.
In this model, moist fuel dries, then the dried fuel thermally decomposes to form char
by two paths: fuel pyrolysis and fuel oxidation. $\alpha$-Char forms via fuel pyrolysis while
$\beta$-char forms from fuel oxidation. Next, $\alpha$- and $\beta$-char oxidize and form ash.
The drying and fuel-pyrolysis reactions are endothermic reactions while the fuel- and
char-oxidation reactions are exothermic. When considering 100\% cellulose (i.e., neat cellulose)
the model contains five global reactions, while for mixtures of cellulose and hemicellulose
the model includes 10 global reactions. The full 10-step chemical kinetic model follows:
\begin{align}
\ce{Cellulose.$v_{w,dr}$H_2O  &-> Cellulose + $v_{w,dr}$H_2O(g)} \reactiontag \\
\ce{Cellulose  &-> $v_{\alpha,cp}$ $\alpha$-Char_{c} + $v_{g,cp}$ Gas} \reactiontag \\
\ce{Cellulose + $v_{O_2,co}$O_{2}  &-> $v_{\beta,co}$ $\beta$-Char_{c} + $v_{g,co}$Gas} \reactiontag \\
\ce{$\alpha$-Char_{c} +$v_{O_2,c\alpha o}$ O_{2} &-> $v_{a,c\alpha o}$Ash_{c} + $v_{g,c\alpha o}$Gas} \reactiontag \\
\ce{$\beta$-Char_{c} + $v_{O_2,c\beta o}$O_{2} &-> $v_{a,c\beta o}$Ash_{c} + $v_{g,c\beta o}$Gas} \reactiontag
\end{align}
\begin{align}
\ce{Hemicellulose.$v_{w,dr}$H_2O  &-> Hemicellulose + $v_{w,dr}$H_2O(g)} \reactiontag \\
\ce{Hemicellulose &-> $v_{\alpha,hp}$ $\alpha$-Char_{h} + $v_{g,hp}$ Gas} \reactiontag \\
\ce{Hemicellulose + $v_{O_2,ho}$ O_{2}  &->  $v_{\beta,ho}$ $\beta$-Char_{h} + $v_{g,ho}$ Gas} \reactiontag \\
\ce{$\alpha$-Char_{h} + $v_{O_2,h\alpha o}$O_{2} &-> $v_{a,h\alpha o}$Ash_{h} + $v_{g,h\alpha o}$Gas} \reactiontag \\
\ce{$\beta$-Char_{h} + $v_{O_2,h\beta o}$O_{2} &-> $v_{a,h\beta o}$Ash_{h} + $v_{g,h\beta o}$Gas} \reactiontag
\end{align}
where $v$ is the stoichiometric coefficient;
$\alpha$ and $\beta$ indicate char produced from fuel pyrolysis and fuel oxidation reactions, respectively;
and subscripts $w$, $g$, $O_2$, $a$, $c$, $h$, $dr$, $o$, $p$, $\alpha o$, $\beta o$ are
water, gas, oxygen, ash, cellulose, hemicellulose, drying, oxidation, pyrolysis,
$\alpha$-char oxidation, and $\beta$-char oxidation, respectively.

Table~\ref{tab:kinetic} lists the chemical-kinetic parameters (pre-exponential factor, 
activation energy, order of reaction, and heat of reaction) for the schemes used here, 
obtained from Huang and Rein~\cite{Huang2016}.
They developed the model to simulate smoldering of biomass, by optimizing kinetic parameters 
to match thermogravimetric-analysis measurements using a genetic algorithm~\cite{Li2014}.
We chose the kinetic parameters based on experiments using low-mineral moss peat (2.1\% inorganic 
content), with oxygen concentrations of \SIlist{0;10;21}{\%} and heating rates of 
\SIlist{10;20;30}{\kelvin\per\minute};
the optimized model showed a minimum error of 5.5\% with respect to the experimental measurements.
Here, we apply this model to simulate smoldering in more-general mixtures of cellulose and 
hemicellulose.
We accounted for the consumption of oxygen using the relation
$\upsilon_{\text{O}_2,k}= \Delta H/(-13.1)$ \si{\mega\joule\per\kilo\gram}~\citep{Hugget1980,Huang2015}.

\begin{table}[htbp]
\centering
\tbl{Kinetic parameters for cellulose and hemicellulose model.}
{\begin{tabular}{@{}l c c c c c c@{}}
\midrule
\multicolumn{7}{c}{Cellulose}\\
  Reaction &Reaction & $\log Z $& $E$  & $\Delta H$  & $n$ & $n_{\ce{O2}}$\\
   number&&$\log\si{s^{-1}}$&$\si{\kilo\joule\per\mole}$&$\si{\mega\joule\per\kilo\gram}$&$-$&$-$\\
\midrule
 (R1) &Drying & 8.12    & 67.8 & 2.26 & 2.37 & $-$\\
 (R2) &Pyrolysis & 11.7    & 156 & 0.5 & 1 & $-$\\
 (R3) &Oxidation & 24.2  & 278   & -28.2 & 1.73 & 0.74\\
 (R4) &$\beta$-char oxidation &7.64 & 120 & -28.8   &1.25   &0.89 \\
 (R5) &$\alpha$-char oxidation & 12.2 & 177 & -27.8& 0.93 &0.52\\
\midrule
\multicolumn{7}{c}{Hemicellulose}\\
  Reaction  &Reaction & $\log Z $& $E$  & $\Delta H$  & $n$ & $n_{\ce{O2}}$\\
  Number&&$\log\si{\s^{-1}}$&$\si{\kilo\joule\per\mole}$&$\si{\mega\joule\per\kilo\gram}$&$-$&$-$\\
\midrule
 (R6) &Drying & 8.12    & 67.8 & 2.26 & 2.37 & $-$\\
 (R7)& Pyrolysis & 6.95 & 93.8 & 0.5 & 0.98 & $-$\\
 (R8)& Oxidation & 20.2 & 294 & -20.9 & 0.47 & 0.11\\
 (R9)& $\beta$-char oxidation & 7.64 & 120 & -28.8 &1.25 & 0.89 \\
 (R10)& $\alpha$-char oxidation & 12.2 & 177 & -27.8 & 0.93 & 0.52\\
\bottomrule
\end{tabular}}
\label{tab:kinetic}
\end{table}

\subsection{Physical properties}
\label{sec:physical_properties}

Table~\ref{physical_1} reports the physical properties of condensed-phase
species: solid density ($\rho_{s,i}$), thermal conductivity ($k_{s,i}$),
and heat capacity ($c_i$). 
For the natural bulk densities of cellulose and hemicellulose ($\rho_i$), 
we used the values experimentally measured by Cowan et al.~\cite{Cowan2017}:
\SI{175}{\kilogram\per\meter^3} and \SI{695.71}{\kilogram\per\meter^3},
respectively.
(Bulk density refers to the density of the species including pores, i.e.,
total mass divided by total volume, while solid density is the density 
of the species without any pores.)
We calculated the bulk density of char using 
the correlation $\rho_{\text{char}} \approx \upsilon_{\text{char}}\times\rho_{\text{fuel}}$ \citep{Huang2017}
and the bulk density of ash using
$\rho_{\text{ash}} \approx \text{AC}/100\times10\times\rho_{\text{fuel}}$,
where AC stands for ash content~\citep{Huang-comm}.
The ash contents of cellulose and hemicellulose are $0.3\%$ and $1.2\%$,
respectively~\citep{Moriana2014,Huang2016,Cell_process}. 
Following the studies of Huang et al.~\cite{Huang2015}, 
we assumed the solid physical properties of fuels do not depend on temperature.

\begin{table}[htbp]
\centering
\tbl{Thermophysical properties of condensed-phase species, 
taken from the literature for water \citep{Huang2016}, cellulose \citep{CTT},
hemicellulose \citep{aseeva2014,Thybring2014,Eitelberger2011},
char \citep{Huang2016,bejan2003heat}, 
and ash \citep{Huang2016,bejan2003heat}.}
{\begin{tabular}{@{}l c c c@{}}
\toprule
Species & Solid density, $\rho_{s,i}$ & Thermal conductivity, $k_{s,i}$ & Heat capacity, $c_i$  \\ 
 & (\si{\kilo\gram\per\meter^3})  & ({\si{\watt\per\meter\per\kelvin}}) & ({\si{\joule\per\kilo\gram\per\kelvin}}) \\
\midrule
Water & 1000 & 0.6 & 4186 \\
Cellulose      & 1500 &  0.356 & 1674 \\
Hemicellulose & 1365 & 0.34 & 1200 \\
Char      & 1300 &  0.26 & 1260 \\
Ash       & 2500 &  1.2 & 880 \\
\bottomrule
\end{tabular}}
\label{physical_1}
\end{table}

The effective thermal conductivity of a condensed-phase species is calculated using
\begin{equation}
  k_i=k_{s,i}(1-\psi_{i})+\gamma_i \sigma T^3 \;,
\end{equation}
where $k_{s,i}$ is the solid thermal conductivity of species $i$, 
$\psi_{i}$ is the porosity of species $i$,
$\sigma$ is the Stefan--Boltzmann constant,
and $\gamma_i$ is an empirical parameter for radiation across pores that depends 
on pore size~\citep{Lautenberger2009}. 
The porosity of species $i$ is calculated with
\begin{equation}
    \psi_{i} = 1 - \frac{\rho_i}{\rho_{s,i}} \;.
\label{eq:porosity}
\end{equation}
Pore size, $\gamma_i$, and permeability are calculated for each condensed-phase species 
at their natural densities using
\begin{align}
d_{po,i} &\approx d_{p,i} = \frac{1}{S_i\times\rho} \label{pore_size_1} \\
K_i &= \num{e-3} \times d_{p,i}^2 \label{permeablity_1} \\
\gamma_i &= 3\times d_{po,i} \label{gamma_1} \;,
\end{align}
where $\rho$ is the density of the fuel, $S_i$ is the particle surface area for species $i$,
$d_{p,i}$ is the particle size, $K_i$ is the permeability, and $d_{po,i}$ is the 
pore size \citep{punmia2005soil,Yu2006,Huang2015,Huang2016}.
The particle surface areas of cellulose, cellulose-based ash, hemicellulose,
and hemicellulose-based ash are \SIlist{0.0388;0.1533;0.0678;0.2712}{\meter^2\per\gram},
respectively~\citep{Sigma-Aldrich,Huang2016,S_ash,parchem}. These correlations apply at the natural
densities of the fuels based on the assumption of similar particle and pore 
size~\citep{Huang2016,Huang2015}. For cases where we model fuels with specific or varying densities,
we assigned this value as the natural density and used Eqs.~\eqref{pore_size_1}--\eqref{gamma_1} 
to vary properties with density.

However, when we emulate increases in density due to compression, the particle size $d_{p,i}$ 
remains constant but pore size $d_{po,i}$ decreases due to the reduction of pore volume.
Thus, when validating our model (Section~\ref{sec:validation}), we used the
experimental measurements of Smucker et al.~\cite{Smucker:2019sm,Smucker2021correction} for bulk density; 
they changed the density of fuels by compressing the samples from their
natural density to reach the desired density. To model this compression,
we account for the associated changes in pore size ($d_{po,i}$)
and radiation parameter ($\gamma_i$) by scaling them with change in porosity ($\psi$), 
since porosity is directly proportional to the volume occupied by pores.
Permeability also changes during compression, which we vary with
the Kozeny--Carman equation:
\begin{equation}
  K_i \propto \frac{e_i^3}{1 + e_i} \;,
\end{equation}
where $e_i$ is the void ratio, related to porosity with $e_i = \psi_i / (1-\psi_i)$.

Unless mentioned otherwise, we ran all simulations with 10\% moisture content to account for
moisture content already present in natural fuels and moisture absorbed from the
atmosphere~\citep{Huang2016,Huang2017}. The addition of water changes the
density of the (wet) fuel, and we accounted for this change using
\begin{equation}
  \rho_{\text{wet fuel}}=\rho_{\text{dry fuel}}\times(1+\text{MC}) \;,
\end{equation}
where $\text{MC}$ is the moisture content~\citep{Huang2015}.
To investigate the role of this natural fuel expansion, we considered cases
where the fuel expands with moisture content and where it does not;
when the fuel does expand, we use the correlation developed for peat by 
Huang and Rein~\cite{Huang2017} with the bulk density modified for the fuels considered
here. Porosity changes less than 5\% with this change in density here, 
so we consider this adoption justified. The modified correlation is
\begin{equation}
  \rho_{\text{dry fuel}} = \frac{200 + 40 \text{MC}}{1 + \text{MC}} \;.
\end{equation}
Thermal conductivity $(k)$ and heat capacity $(c)$ also vary with moisture content, and we change
those for wet fuels by averaging using volume fraction ($X_{i}$) and
mass fraction ($Y_{i}$), respectively~\citep{Huang2015,Lautenberger2009}:
\begin{align}
  k_{\text{wet fuel}} &= X_{\ce{H2O}} k_{\ce{H2O}} + X_{\text{dry fuel}} k_{\text{dry fuel}} \\
  c_{\text{wet fuel}} &= Y_{\ce{H2O}} c_{\ce{H2O}} + Y_{\text{dry fuel}} c_{\text{dry fuel}} \;.
\end{align}

\subsection{Calculation of global quantities}

\begin{figure}[htbp]
\centering
\includegraphics[width= 0.8\linewidth]{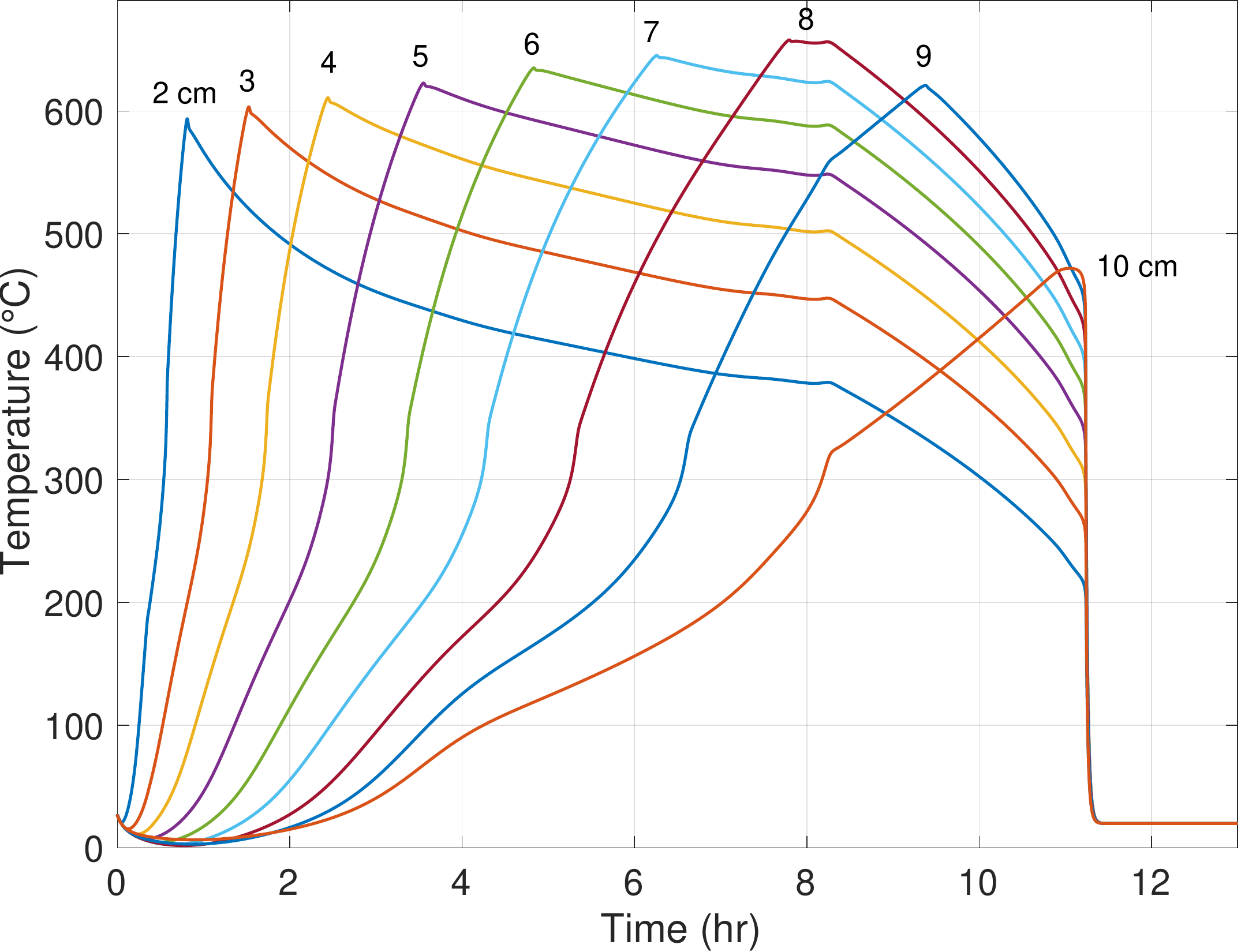}
\caption{Temperature profile with respect to time for a fuel composition of \SIlist{50}{\%} cellulose at a density of
\SIlist{300}{\kilogram\per\meter^3}.}
\label{fig:temperature_profiles_example}
\end{figure}

The two main parameters of interest in this study are mean propagation speed of smoldering and
mean peak temperature. We calculate propagation speed by numerically computing the derivative of
depth with respect to time of peak temperature; in other words, the difference between two depths
divided by the times when those depths reach their maximum temperature.
Figure~\ref{fig:temperature_profiles_example} shows an example temperature vs.\ time profile that 
demonstrates how we record the data for calculating these global quantities.
This requires selecting a depth interval for evaluating this finite difference; to determine
the appropriate interval value for calculating mean propagation speed, starting at \SI{6}{\centi\meter}
we systematically reduced the depth interval and examined the effect on calculated mean propagation speed.
(A smaller depth interval requires both producing and evaluating more data from the simulations, so we
seek a pragmatic choice that affects the results little while reducing the computational burden.)
After reducing the depth interval to \SI{1}{\centi\meter}, further reduction negligibly affects
propagation speed: reducing from \SI{1}{\centi\meter} to \SI{0.5}{\centi\meter} increases the
calculated speed by less than 0.3\%. 
As a result, we chose a depth interval of \SI{1}{\centi\meter} for all cases.
The supplementary material shows the effects of reducing depth interval on propagation speed
in more detail. 
We calculated mean peak temperature similarly by averaging the peak
temperatures every \SI{1}{\centi\meter}.

\section{Results and discussion}
\label{sec:results}

First, we validated the model by comparing it with experimental measurements of mean propagation
speed and mean peak temperature. Then, we varied density and fuel composition to study how these
parameters affect smoldering behavior. Next, we examined the effect of moisture content
on mean peak temperature and mean propagation speed for 100\% cellulose. Finally, we
examined how the critical moisture content of ignition and extinction change with fuel composition.

\subsection{Validation}
\label{sec:validation}

We validated the computational model using the experimental results of
Smucker et al.~\cite{Smucker:2019sm,Smucker2021correction} by comparing two parameters:
mean peak temperature and mean propagation speed. The experiments used a one-dimensional
reactor box of dimensions $\SI{10}{\cm} \times \SI{10}{\cm} \times \SI{13}{\cm}$, with thermocouples placed
at $\SI{1}{\cm}$ depth intervals. The top surface of the reactor box was open to atmosphere
and the other sides were insulated using a calcium silicate insulation board, 
and the fuel samples were ignited using a cartridge heater applied until the point of 
self-sustained smoldering.
The supplemental material contains key information about the experimental measurements, and
Smucker et al.~\cite{Smucker:2019sm,Smucker2021correction} provide further details about the 
experimental configuration.

To ensure self-sustained smoldering, we performed our validation simulations using a heat flux of
\SI{25}{\kilo\watt\per\meter^2} applied for \SI{20}{\minute} at the top surface.
However, we found that smoldering behavior was insensitive to the magnitude
of the heat flux; doubling it changed the propagation speed by less than 1.8\%.
This gave us the confidence to use heat flux as the boundary condition to ignite the fuel sample.

\begin{figure}[htbp]
\centering
\includegraphics[width= 0.9\linewidth]{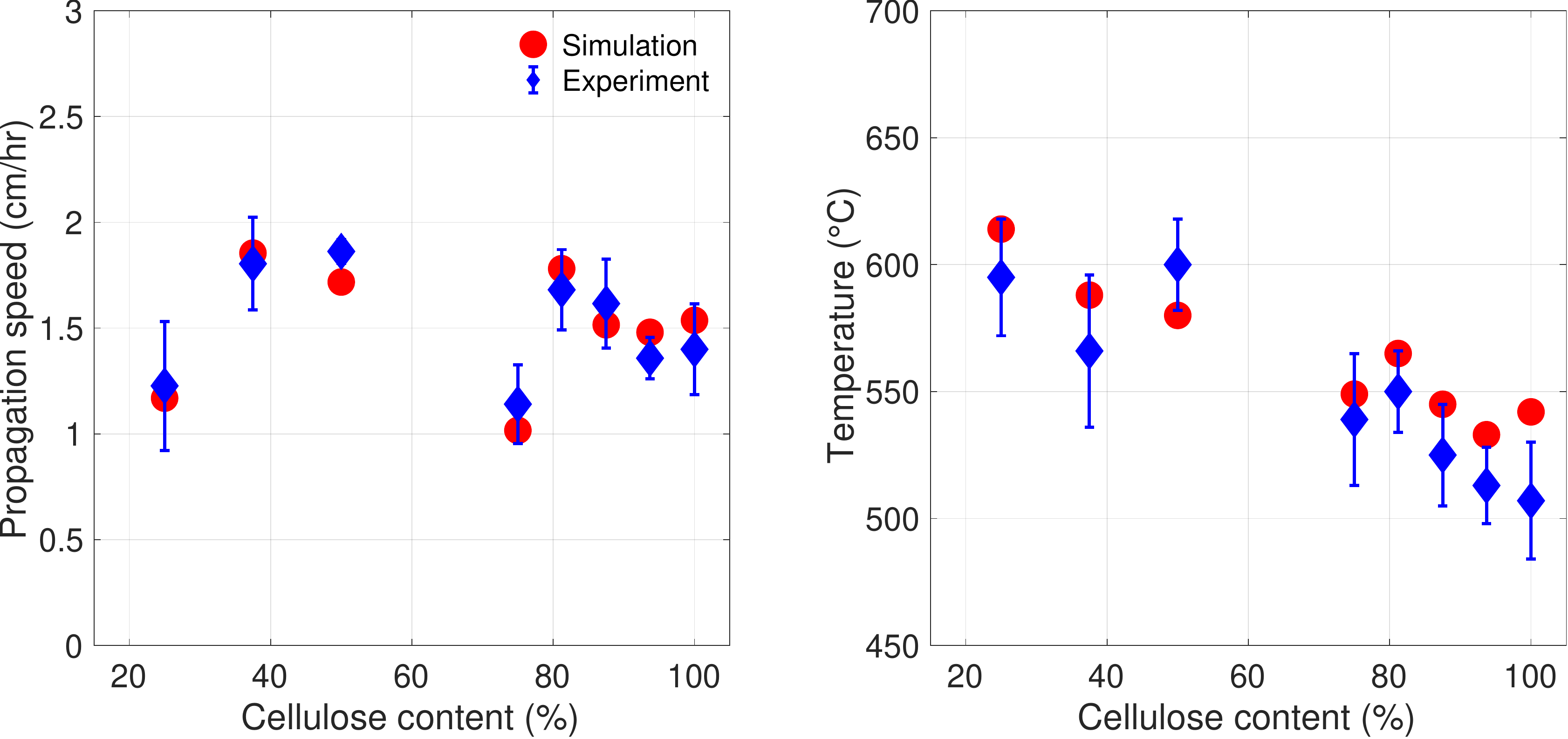}
\caption{Experimental (diamond) and predicted (circle) propagation speeds and mean peak temperatures
(filled symbols) for fuel compositions of \SIlist{25;50;75;100}{\%} cellulose at densities of
\SIlist{400;250;300;170}{\kilogram\per\meter^3}.}
\label{fig:validaton}
\end{figure}

We used eight fuel samples with varying fuel composition and density to validate the model:
\SIlist{25;37.5;50;75;81.2;87.5;93.7;100}{\%} cellulose, with the remainder hemicellulose, 
at respective densities of \SIlist{400;200;250;300;250;250;250;170}{\kilogram\per\meter^3}.
They created these mixtures artificially by mixing the two components and then 
compressing the fuel to achieve a desired density.
Figure~\ref{fig:validaton} compares the experimental
measurements and model calculations of mean propagation speed and peak temperature;
the model captures all of the experimentally observed trends, and also agrees well quantitatively.
The model overpredicts mean propagation speed for 100\% cellulose by 10.8\% in the worst case,
while the average error in propagation speed for the four mixtures is 8.7\%. Similarly, the
model overpredicts mean peak temperature for 100\% cellulose in the worst case by 6.1\%,
with the average error at 5.3\%.
Based on these results, we will use this model for the remaining
studies here.

\subsection{Sensitivity to fuel composition and density}
\label{sec:composition-density}

\begin{figure}[htbp]
\centering
\includegraphics[width=0.9\linewidth]{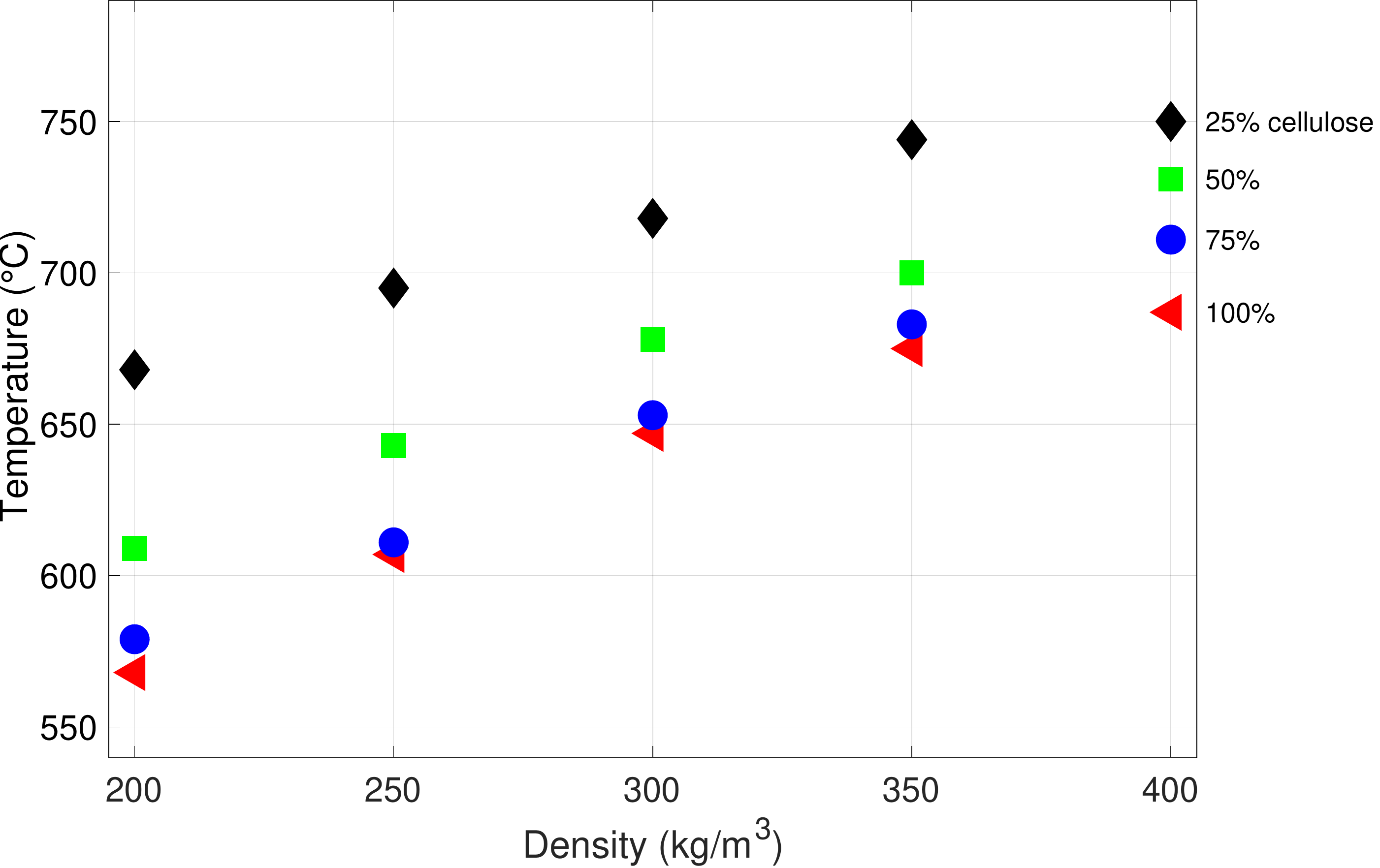}
\caption{Effects of varying density and fuel composition on peak temperature.}
\label{temp_comp_dens}
\end{figure}

\begin{figure}[htbp]
\centering
\includegraphics[width=0.9\linewidth]{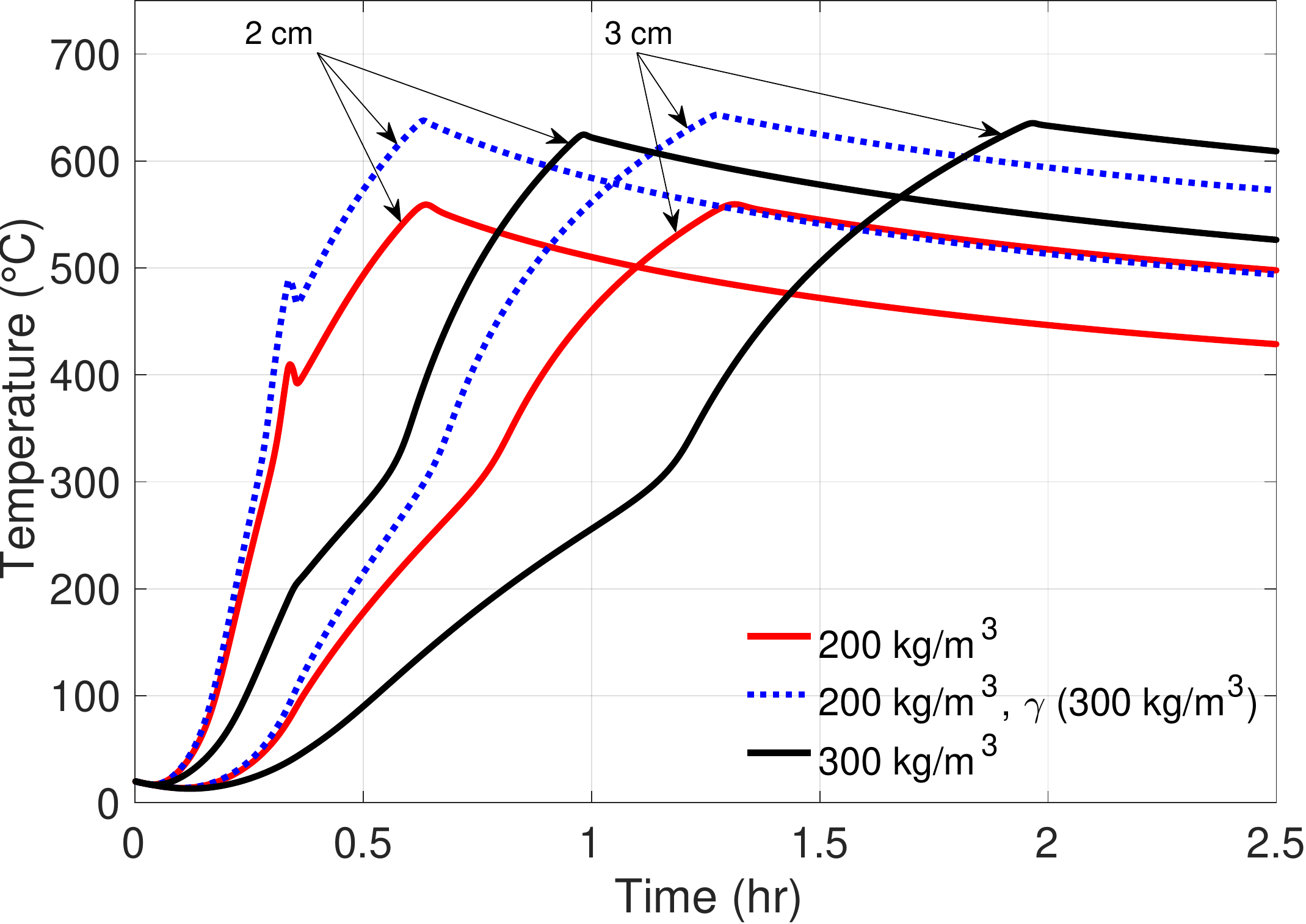}
\caption{Temperature profiles at depths of \SIlist{2;3}{\centi\meter} of 100\% cellulose with densities \SI{200}{\kilogram\per\meter^3}, \SI{300}{\kilogram\per\meter^3}, and \SI{200}{\kilogram\per\meter^3} with the empirical parameter for radiation across pores ($\gamma$) of \SI{300}{\kilogram\per\meter^3}.}
\label{temp_k_all}
\end{figure}

Next, we investigated the effects of density and fuel composition on mean peak temperatures,
as Figure~\ref{temp_comp_dens} shows.
We artificially created these mixtures to analyze the effects of fuel composition and density on smoldering behavior.
We varied fuel density between \SIrange{200}{400}{\kilogram\per\meter^3} in increments of
\SI{50}{\kilogram\per\meter^3} and the fuel composition from 100--25\% cellulose in decrements of 25\% cellulose,
with hemicellulose as the remaining fuel in the mixture.
As Figure~\ref{temp_comp_dens} shows, mean peak temperature increases with increasing 
density.\footnote{The calculated peak temperatures differ from those shown in 
Figure~\ref{fig:validaton} due to the different treatment of density, as we discussed in 
Section~\ref{sec:physical_properties}.}
To determine the cause of this temperature dependence, 
we individually varied the parameters
that change when density increases.
We found that decreasing value of the empirical parameter for radiation
across pores $(\gamma)$ of the condensed-phase species contributes most to the increase in peak temperatures.
Figure~\ref{temp_k_all} shows temperature profiles for 100\% cellulose at densities of
\SIlist{200;300;200}{\kilogram\per\meter^3} but with $\gamma$
associated with \SI{300}{\kilogram\per\meter^3}.
For the fuel with a density of \SI{200}{\kilogram\per\meter^3},
when we change only the values of $\gamma$ for the condensed-phase species to those at
\SI{300}{\kilogram\per\meter^3}, the peak temperatures closely match those of the \SI{300}{\kilogram\per\meter^3} fuel, 
with differences of 1.9\% and 1.2\% for \SI{2}{\centi\meter} and \SI{3}{\centi\meter} 
profiles, respectively.

\begin{figure}[htbp]
\centering
\includegraphics[width= 0.9\linewidth]{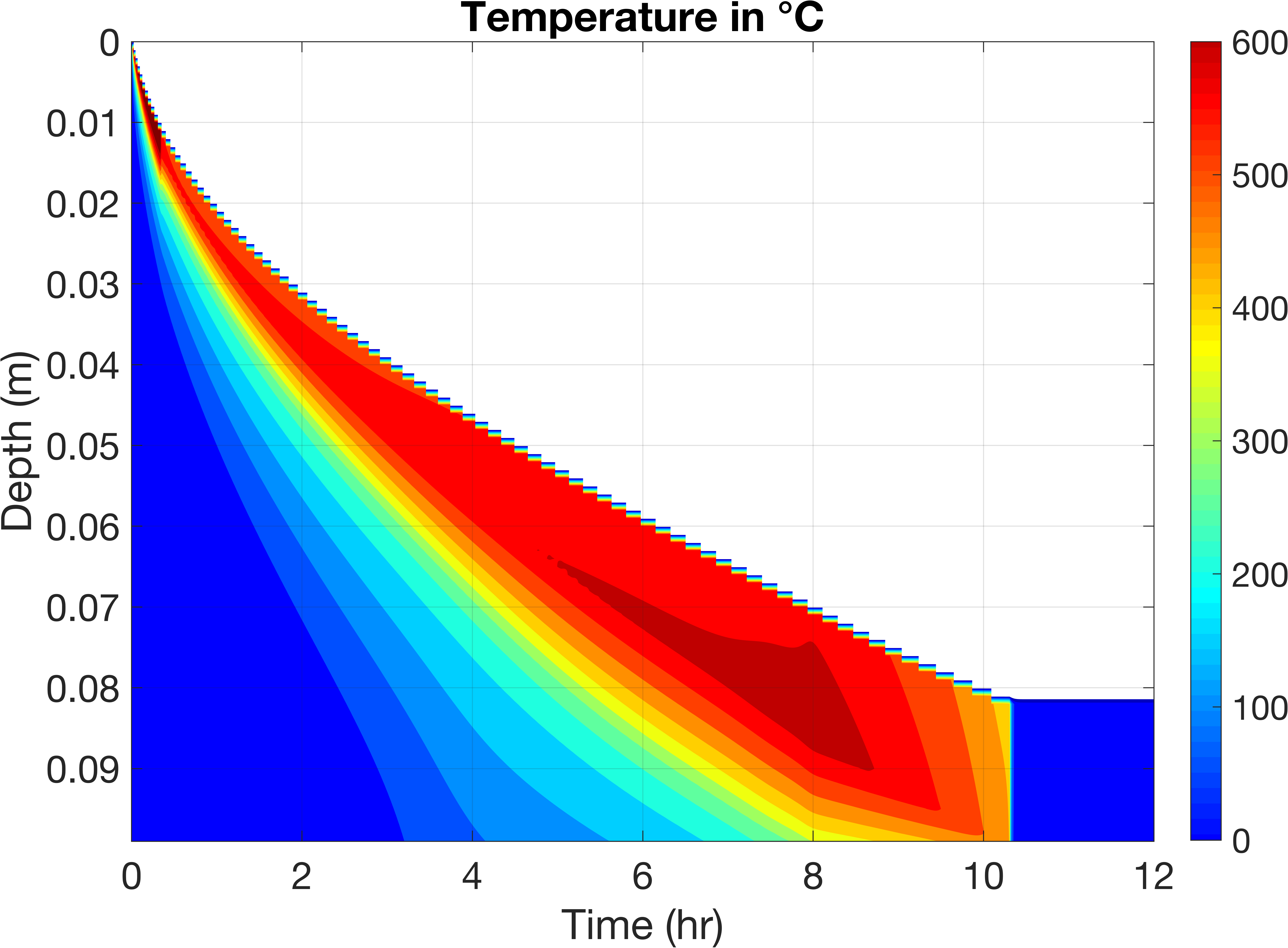}
\caption{Temperature contour varying with depth and time for fuel composition of cellulose 50\% and density of \SI{300}{\kilogram\per\meter^3}}
\label{temp_cont}
\end{figure}

Figure~\ref{temp_comp_dens} also shows that increasing hemicellulose content in the fuel increases mean peak temperature.
To explain this, Figure~\ref{temp_cont} shows temperature at varying depths and times for 50\% cellulose and
50\% hemicellulose at a density of \SI{300}{\kilogram\per\meter^3}. 
The peak temperatures in Fig.~\ref{temp_cont} do not occur at the surface of the fuel where oxygen is most available, but instead below the surface.
Ash forms at the topmost layer of the fuel, acting as an insulator. 
According to Eq.~\eqref{pore_size_1} and~\eqref{gamma_1}, ash formed from cellulose 
has a higher $\gamma$ than ash formed from hemicellulose.
This leads to greater losses due to radiation across the pores at higher cellulose 
content, hence peak temperatures drop with increasing cellulose content.

\begin{figure}[htbp]
\centering
\includegraphics[width= 0.9\linewidth]{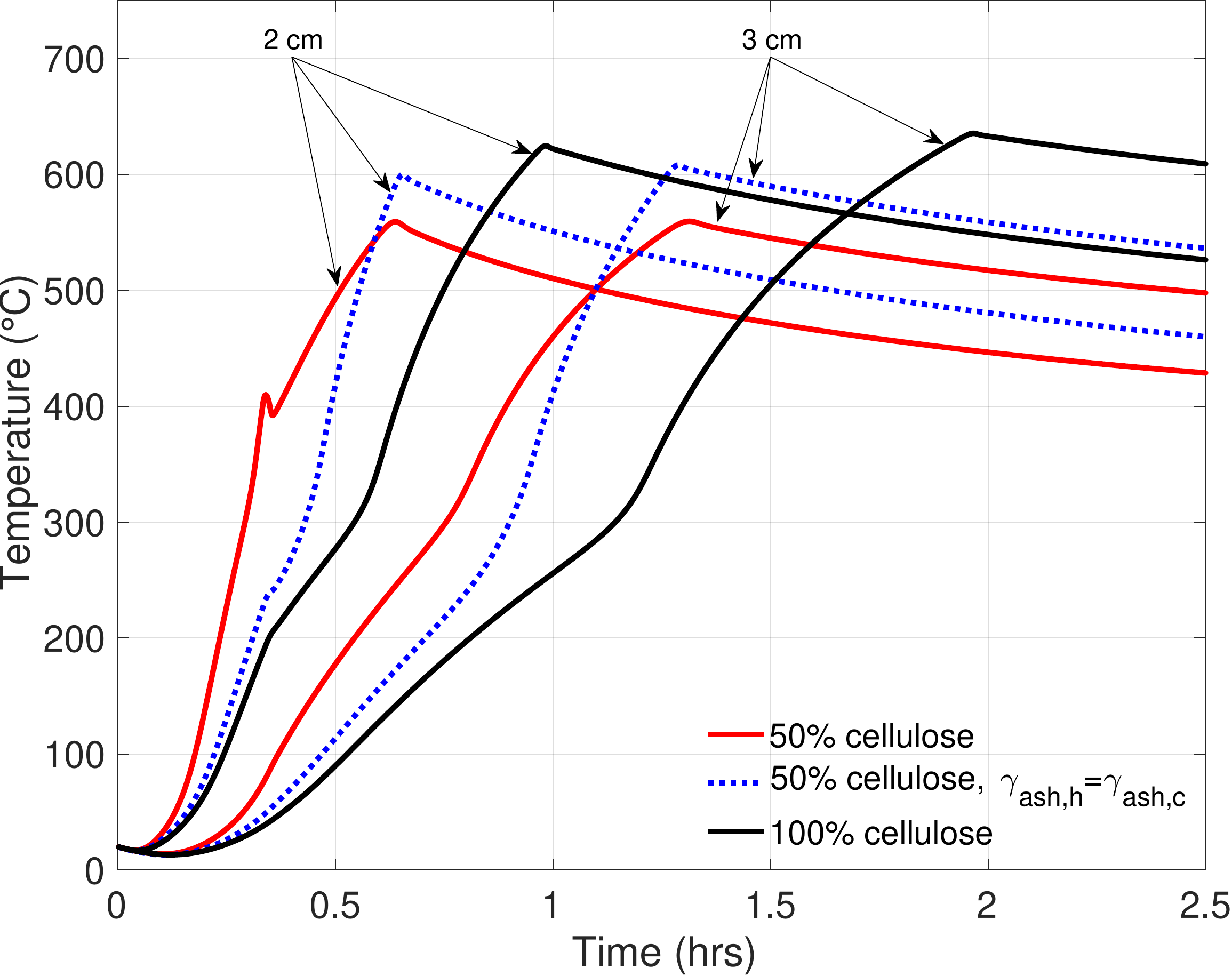}
\caption{Temperature profiles at depth \SIlist{2;3}{\centi\meter} for fuels with density \SI{300}{\kilogram\per\meter^3} and fuel composition of 50\% cellulose, 100\% cellulose, and 50\% cellulose with the parameter for radiation across pores of ash coming from hemicellulose set equal to that from cellulose ($\gamma_{\text{ash,h}} = \gamma_{\text{ash,c}}$).}
\label{temp_k_ash}
\end{figure}

To test this theory, we ran a simulation with the value of $\gamma$ of 
ash from hemicellulose set equal to the $\gamma$ of ash from cellulose for a 
fuel mixture with 50\% cellulose. In other words, in this case the ash 
formed from hemicellulose matches that from cellulose, in terms of radiation
heat transfer across the pores.
Figure~\ref{temp_k_ash} shows the resulting temperature profiles along with temperature 
profiles of 50\% cellulose and 100\% cellulose at density \SI{300}{\kilogram\per\meter^3}.
The peak temperature of 50\% cellulose matches that of 100\% cellulose when $\gamma$ from 
hemicellulose matches that of ash from pure cellulose, with differences of 3.9\% and 4.4\% 
for \SI{2}{\centi\meter} and \SI{3}{\centi\meter} profiles, respectively.

%
Our findings show that the physical parameters of condensed-phase species control 
the observed variations in peak temperature, both as density and fuel composition change.
Richter et al.~\cite{richter2019effect} also discussed the larger role that physical
properties play in wood charring, compared with reaction kinetics. 
Charring, which occurs through pyrolysis and 
heterogeneous oxidation, controls burning behavior and relates to 
temperature profile (including peak temperature).
Figures~\ref{temp_k_all} and~\ref{temp_k_ash} also show that the location of peak temperature 
does not shift significantly even as its value increases,
In Figures~\ref{temp_k_all} the shift is 0.6\% and 2.3\% for \SIlist{2;3}{\centi\meter} 
profiles, and in Figures~\ref{temp_k_ash} the shift is 3.1\% and 2.3\% for 
\SIlist{2;3}{\centi\meter} profiles, respectively.
This indicates that the change in peak temperature does not notably affect propagation speed.

\begin{figure}[htbp]
\centering
\includegraphics[width=0.9\linewidth]{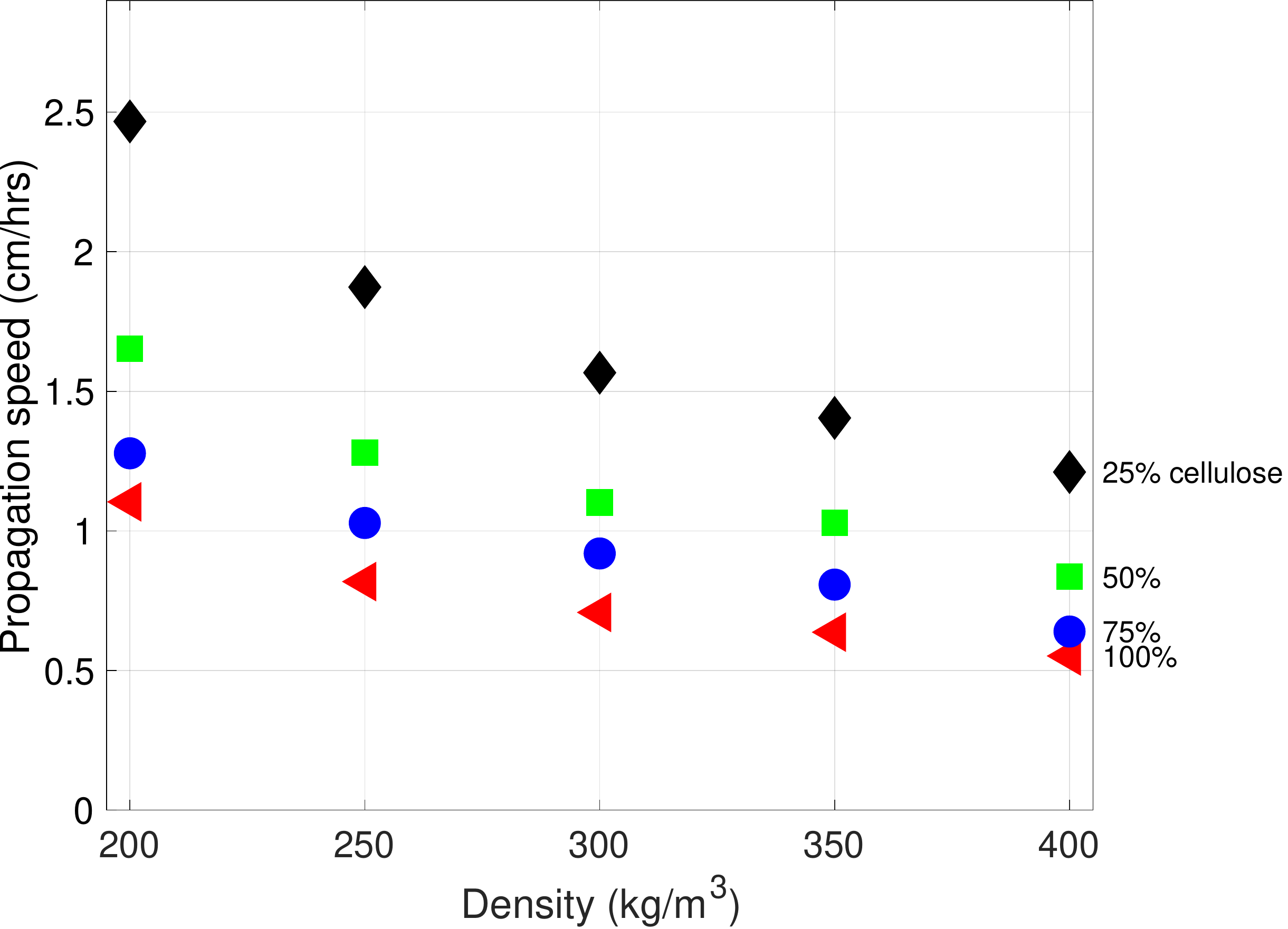}
\caption{Effects of varying density and fuel composition on propagation speed.}
\label{speed_comp_dens}
\end{figure}

\begin{figure}[htbp]
\centering
\includegraphics[width= 0.8\linewidth]{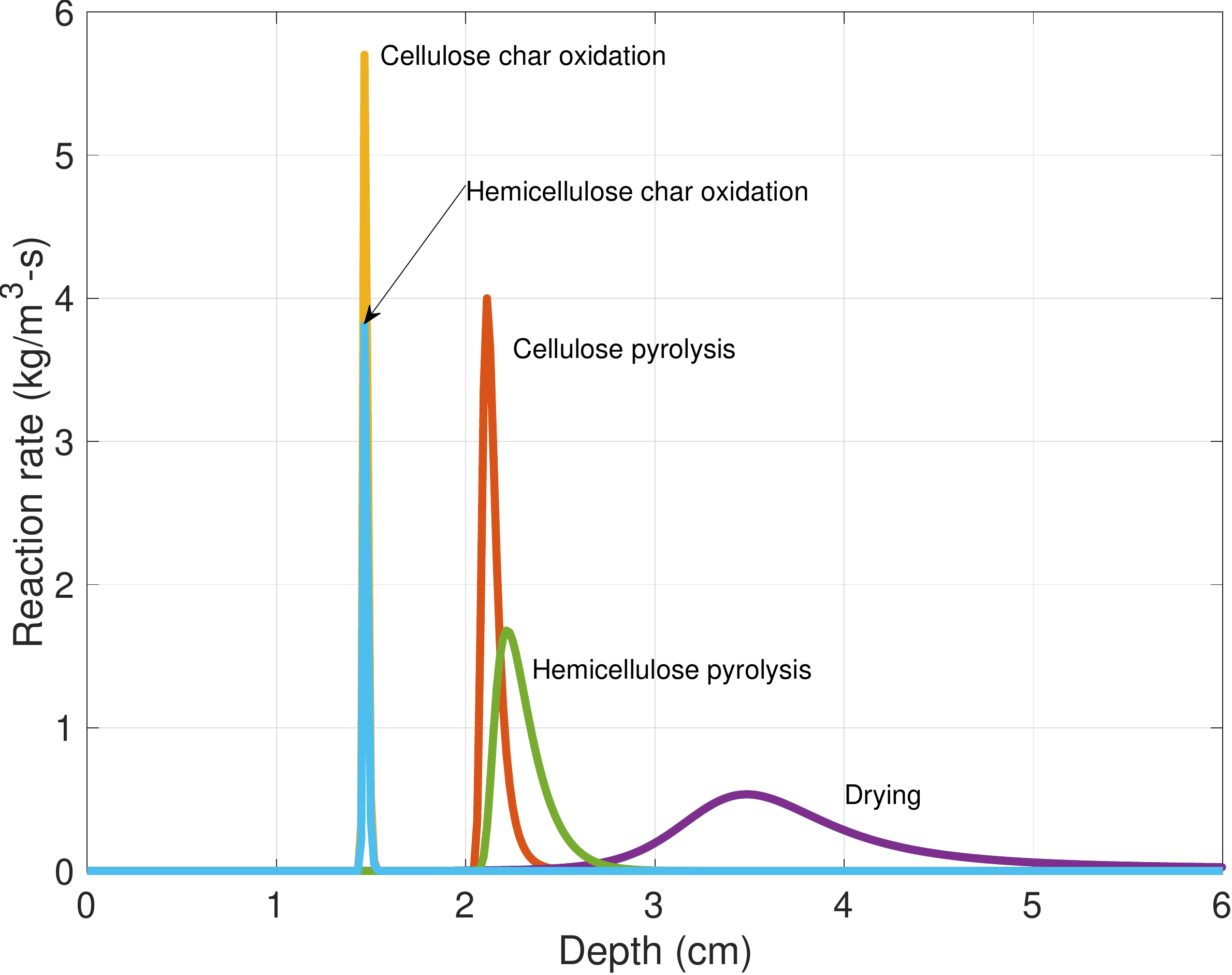}
\caption{Reaction rates of cellulose and hemicellulose drying, pyrolysis, and oxidation 
for 50\% cellulose and 50\% hemicellulose at density \SI{300}{\kilo\gram\per\meter^3} 
along the depth at \SI{4000}{\second}.}
\label{rr_cell_hemi}
\end{figure}

\begin{figure}[htbp]
\centering
\includegraphics[width=\linewidth]{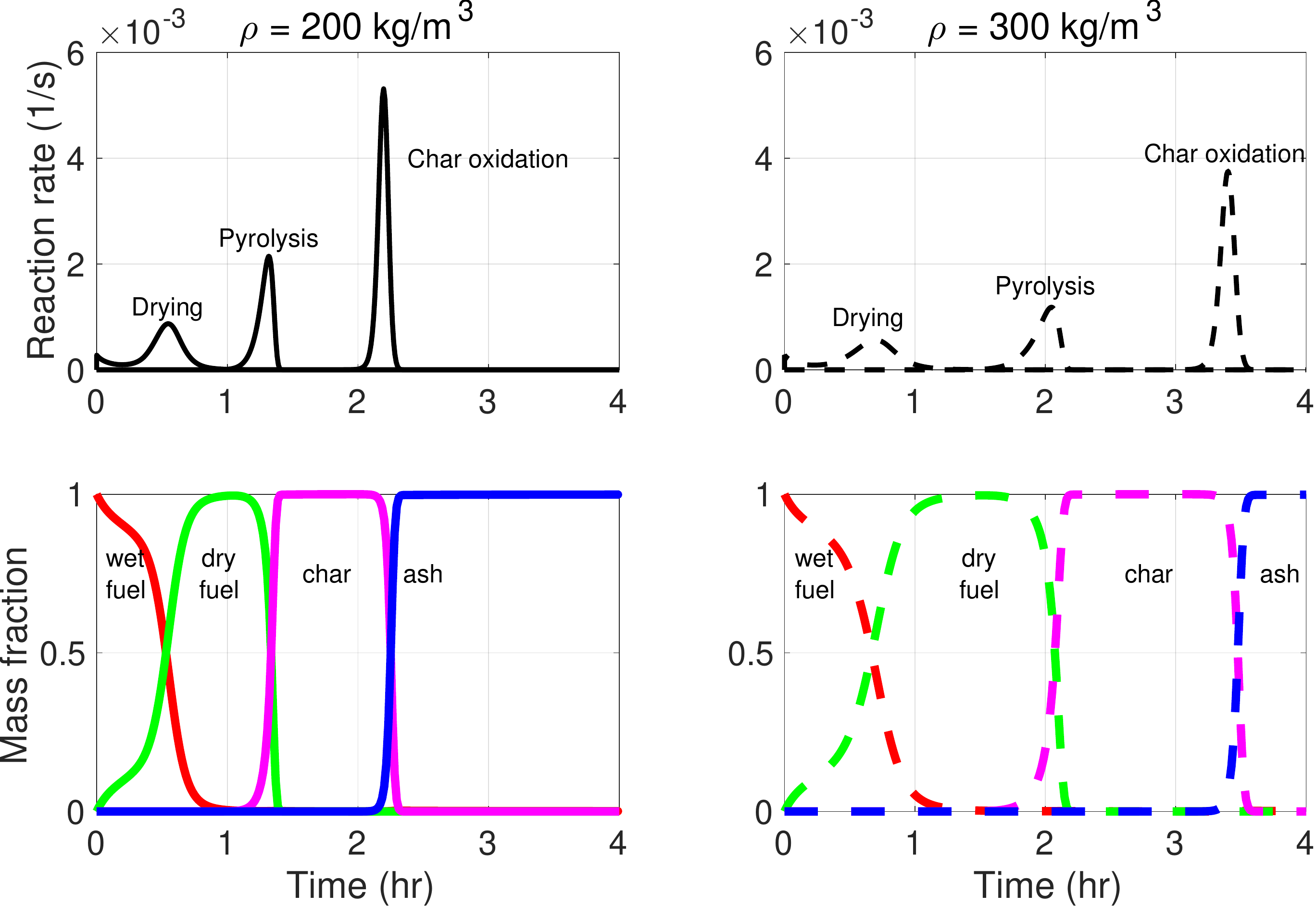}
\caption{Reaction rates of drying, pyrolysis, and char oxidation (top) and mass fractions of wet fuel, dry fuel, char, and ash (bottom) of 100 \% cellulose with density 
\SIlist{200;300}{\kilogram\per\meter^3}.}
\label{rr_and_y_D}
\end{figure}

Next, we consider the effects of density and fuel composition on mean propagation speed, 
shown in Figure~\ref{speed_comp_dens}. Propagation speed increases with increasing  
hemicellulose content and decreases with increasing density. To understand the role of 
fuel composition, Figure~\ref{rr_cell_hemi} shows reaction rates of fuel with hemicellulose 
along the depth at \SI{4000}{\second}.
Hemicellulose pyrolyzes faster than cellulose, so that a given time its pyrolysis occurs 
deeper than fuels with a higher proportion of cellulose.
The fuel shrinks faster, providing earlier access to oxygen ultimately leading into 
faster propagation speed. To examine the role of density, Figure~\ref{rr_and_y_D} shows 
the reaction rates and condensed-phase species mass fractions at \SI{4}{\centi\meter} 
below the surface for 100\% cellulose at densities of \SI{200}{\kilogram\per\meter^3} 
and \SI{300}{\kilogram\per\meter^3}. The reaction rates of lower density fuel are higher 
and less spaced out compared to higher density fuel. This means more time is required 
for fuel to convert to char and ash as observed in the mass fractions of Fig.~\ref{rr_and_y_D}.
This comes from the increased density of the fuel, which means more mass in a given 
volume converts to char and ash. 
As a result, the fuel shrinks, delaying access to oxygen for the char formed.

\begin{figure}[htbp]
\centering
\includegraphics[width= 0.9\linewidth]{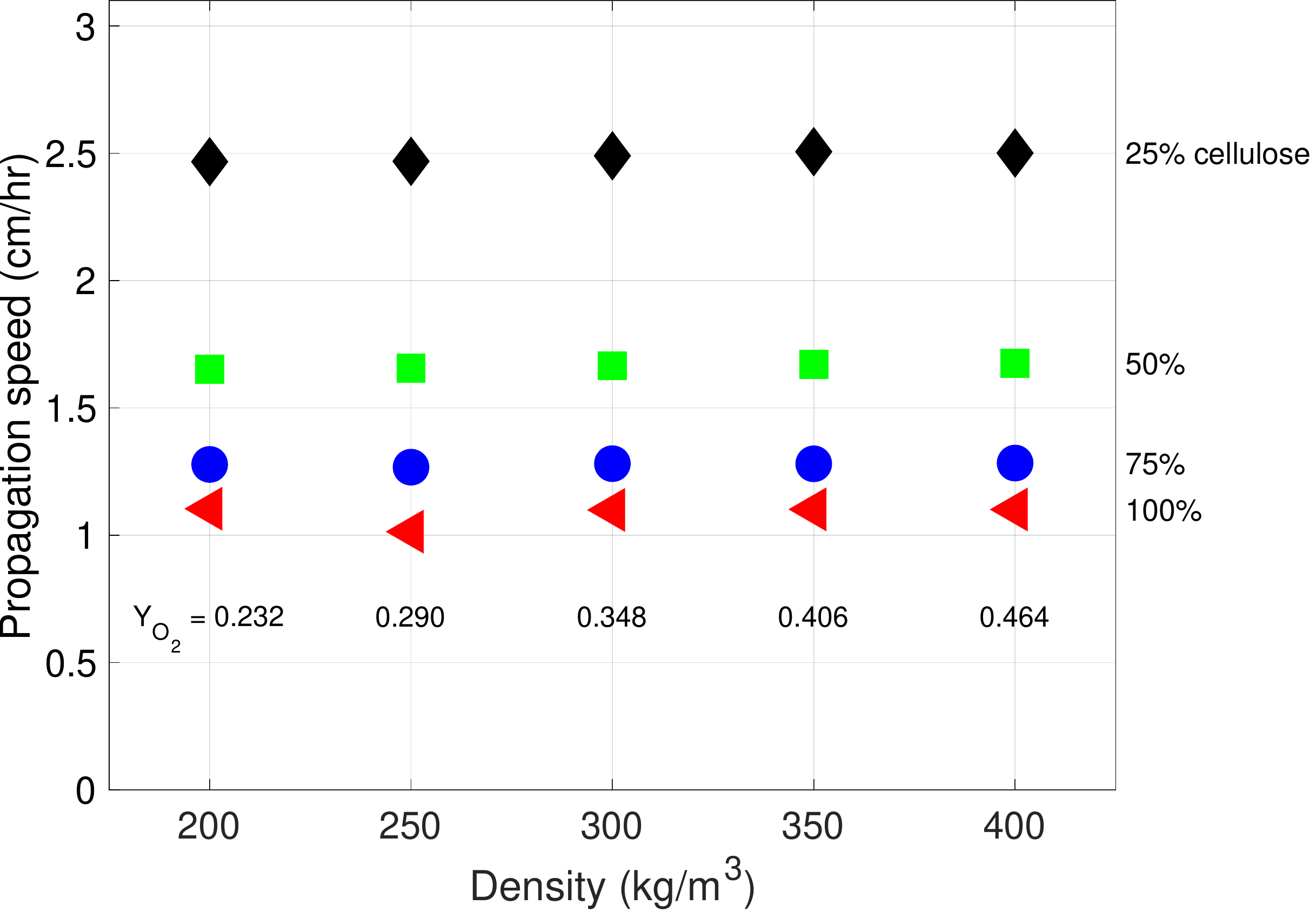}
\caption{Propagation speed when oxygen availability is linearly increased with density, where the value of $Y_{\textrm{O}_2}$ indicates the value of mass fraction of oxygen used for the respective density.}
\label{YJ_speed}
\end{figure}

\begin{figure}[htbp]
\centering
\includegraphics[width= 0.9\linewidth]{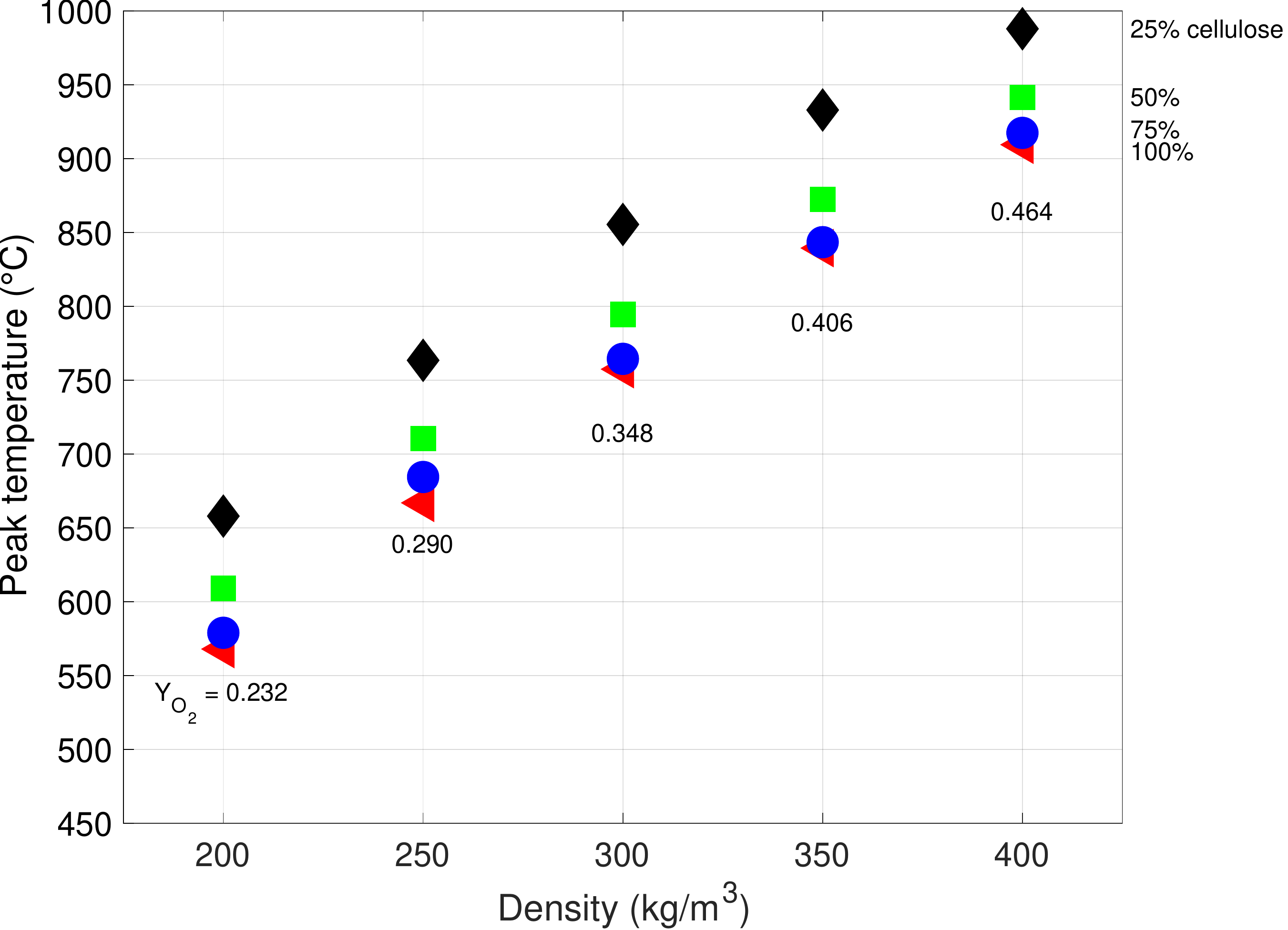}
\caption{Peak temperatures when oxygen availability is linearly increased with density, where the value of $Y_{\textrm{O}_2}$ indicates the value of mass fraction of oxygen used for the respective density.}
\label{YJ_temp}
\end{figure}

Across all fuel compositions, the propagation speed decreases by a factor of two when 
the density of fuel increases proportionally from \SI{200}{\kilogram\per\meter^3} to \SI{400}{\kilogram\per\meter^3} (i.e., doubles).
Huang and Rein~\cite{Huang2017} discussed an inverse relation between oxygen concentration and density.
To further examine the dependence of oxygen concentration, we increased oxygen concentration, 
and density simultaneously, by the same factor.
The oxygen supply was increased via mass fraction of (diffusing) oxygen.
For example, if the density increases by a factor of 1.5, from \SI{200}{\kilogram\per\meter^3} 
to \SI{300}{\kilogram\per\meter^3}, then oxygen mass fraction increased to 0.348 for \SI{300}{\kilogram\per\meter^3}. This was done for all densities and fuel compositions shown in Fig.~\ref{speed_comp_dens}.
Figure~\ref{YJ_speed} shows that when mass fraction of oxygen ($Y_{\ce{O2}}$) 
increases by the same factor as density $(\rho)$ the propagation velocities 
$(S)$ remains constant, confirming the $S \propto Y_{\ce{O2}}/\rho$ relationship posed by 
Huang and Rein~\cite{Huang2017}.
We performed a similar analysis of increasing oxygen supply with density for peak temperatures, 
shown in Figure~\ref{YJ_temp}; peak temperature increases with oxygen content.

To model how propagation speed and peak temperatures quantitatively scale with all the controlling variables, we performed linear regression of the data shown in Figs.~\ref{temp_comp_dens},~\ref{speed_comp_dens},~\ref{YJ_speed}, and~\ref{YJ_temp}.
We used the Matlab function \texttt{regress()}, where the independent variables are mass fraction of cellulose ($Y_{\text{cellulose}}$), density ($\rho$), and oxygen concentration ($Y_{\ce{O2}}$) and the dependent variables are velocity ($S$) and peak temperature ($T$). The resulting equations are:
\begin{align}
S &= 685.08 \times \frac{Y_{\ce{O2}}^{0.9892}}{\rho^{0.9464} \times Y_{\text{cellulose}}^{0.5865}} \; \text{ and}
\label{speed_eqn} \\
T &= 272.28 \times \frac{\rho^{0.2500} \times Y_{\ce{O2}}^{0.3921}}{Y_{\text{cellulose}}^{0.0835}} \;.
\label{temp_eqn}
\end{align}
The goodness of fit ($R^2$) values for both equations are approximately 0.99.
In the fit for propagation speed, Eq.~\eqref{speed_eqn}, the power of fuel density ($\rho$) and mass fraction of oxygen ($Y_{\ce{O2}}$) are 0.9464 and 0.9892, respectively, which are both close to 1.0---confirming the $S \propto \frac{Y_{\ce{O2}}}{\rho}$ relationship discussed earlier.
As demonstrated by Eq.~\eqref{temp_eqn}, peak temperature is more sensitive to oxygen supply than to density.

\subsection{Effect of moisture content on propagation speed}
\label{sec:moisturecontent}

Next, we look into how moisture content affects the propagation speed and peak temperatures 
of smoldering, considering cases both with and without the natural expansion with water.
We investigated cases without expansion because while some prior studies of peat reported 
expansion with addition of water~\citep{Huang2017}, most woody fuels have no reported expansion.

\begin{figure}[htbp]
\centering
\includegraphics[width= 0.9\linewidth]{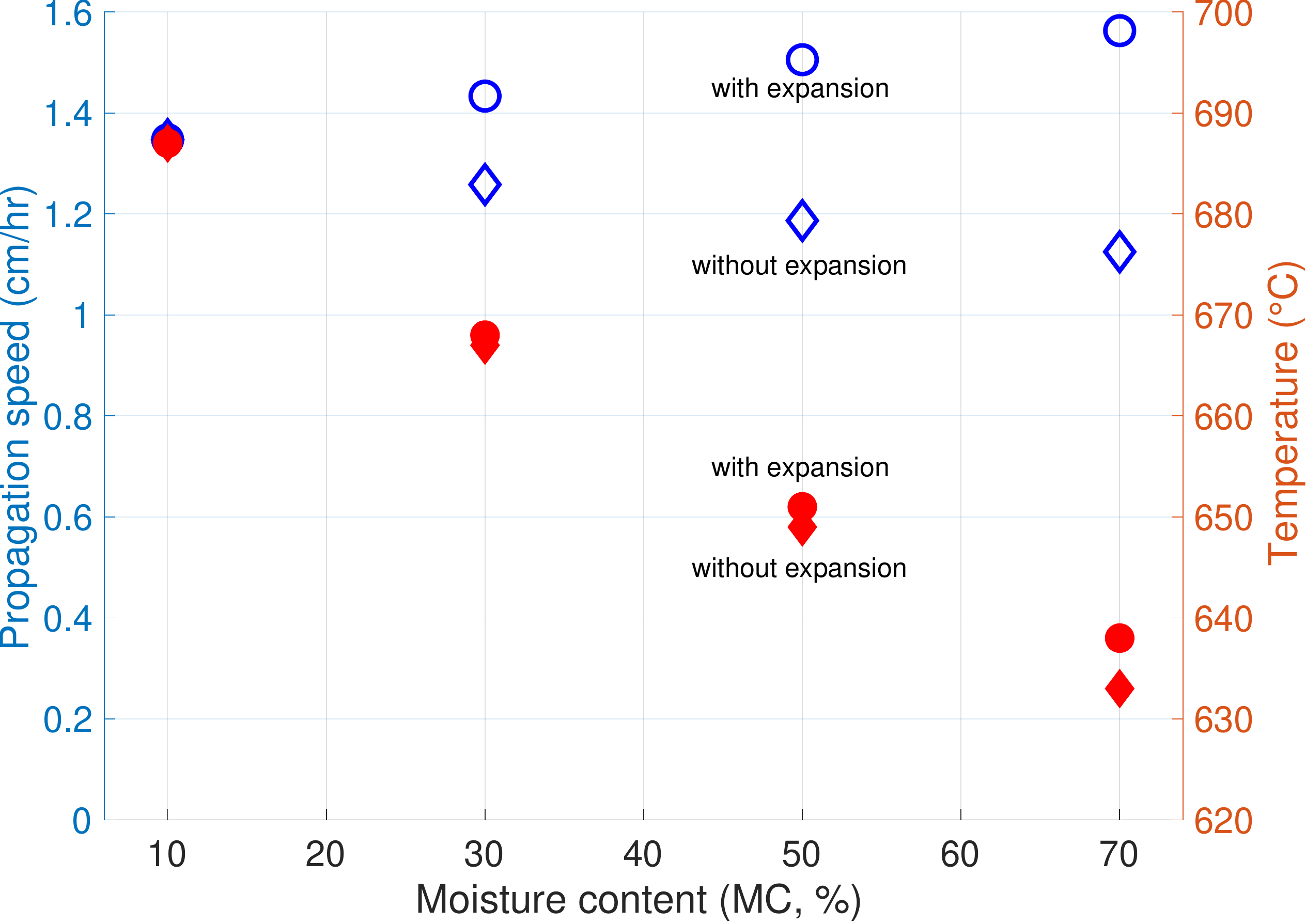}
\caption{Effect of moisture content on propagation speed and peak temperature for 
100\% cellulose, with and without expansion, where the empty symbols indicate 
propagation speed and filled symbols indicate temperature.}
\label{mositure_content_C_100}
\end{figure}

Figure~\ref{mositure_content_C_100} shows the effect of increasing moisture content on 
propagation speed and peak temperature in expanding and non-expanding fuels.
Moisture content is increased from 10\% to 70\% in increments of 20\% for 100\% cellulose.
Peak temperature drops with increasing moisture content with and without expansion, 
while propagation speed shows opposite trends: increasing with expansion and decreasing 
without expansion.

When the fuel does not expand, i.e., all the water added to the fuel sample occupies 
the pores, the propagation speed decreases with increasing moisture content.
In contrast, when the fuel expands, i.e., addition of water increases the total volume
of the fuel, propagation speed increases with moisture content.

\begin{figure}[htbp]
\centering
\includegraphics[width= 0.9\linewidth]{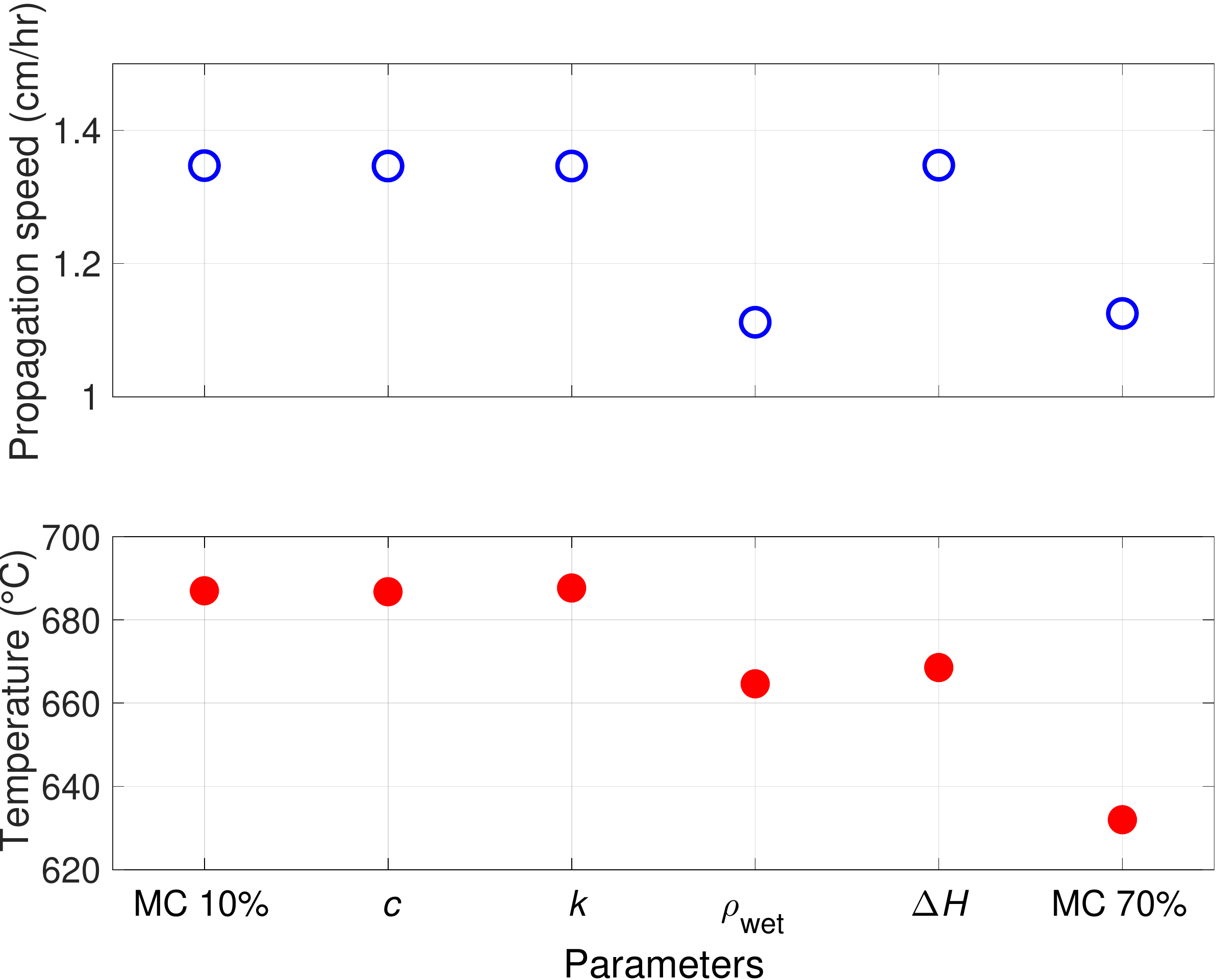}
\caption{Parameter analysis for moisture content without expansion, showing impact of 
parameters on propagation speed (top) and peak temperature (bottom). 
Each parameter ($c$, $k$, $\rho_{\text{wet}}$, and $\Delta H$) was changed to its value 
for 70\% moisture content while holding all other properties to their values at 10\%.
These parameters increased by 55\%, 48\%, 58\%, and 62\%, respectively.
The fully 10\% and 70\% MC cases are shown at the far left and right for comparison; 
note the axis scaling.}
\label{sens_wo_exp}
\end{figure}

Without expansion, when water is added to the fuel the thermal conductivity, heat capacity, 
and wet fuel bulk density increase.
In addition, when moisture content of the fuel increases, the drying becomes more endothermic, 
which increases the associated heat of reaction.
To examine which parameters contribute most to reduce speed and temperature with moisture 
content, we analyzed the affect of each parameter individually as shown in Fig.~\ref{sens_wo_exp}.
To do this, we set the value of each parameter that changes on addition of moisture 
content to the value for 70\% moisture content, keeping all other parameters constant.
The changes in thermal conductivity and heat capacity between the two moisture contents
minimally affect both propagation speed and peak temperature.
Instead, the increase of (wet) bulk density is the main reason for the 
drop in propagation speed. In contrast, the increase in both wet bulk density and 
heat of reaction contribute to the drop in temperature. When increasing the wet 
fuel bulk density, more fuel needs to be dried by the smoldering front in a given volume, 
which decreases both temperature and propagation speed.
When the drying reaction becomes more endothermic, more heat is required to dry the fuel, 
which reduces the peak temperature attained.

When the fuel expands with water addition, the increase in speed could be due to either the 
expansion of the fuel, which reduces density, or the increase in thermal conductivity.
By testing the effect of each parameter, we found that changing only the thermal conductivity 
of the fuel negligibly impacts the propagation speed and temperature, while expansion alone 
increases propagation speed. When a fuel expands the overall density of the fuel decreases, and as
Fig.~\ref{speed_comp_dens} shows when the density of the fuel drops the propagation speed increases.
So, in this case, propagation speed is more influenced by the overall reduction in density 
than the increase in the wet mass of the fuel, which increases the propagation speed.
This result further confirms the relationship Huang and Rein~\cite{Huang2017} first showed for peat.
The temperature reduction in this case comes from the increasing mass of wet fuel and 
increasing endothermicity, similar to the case without expansion.
The temperature trends are similar in both cases, since, as Eq.~\eqref{temp_eqn} shows, 
temperature is comparatively less sensitive to density and thus expansion.

\subsection{Effect of changing composition on critical moisture content}
\label{sec:critical-mc}

\begin{figure}[htbp]
\centering
\includegraphics[width= 0.4\linewidth]{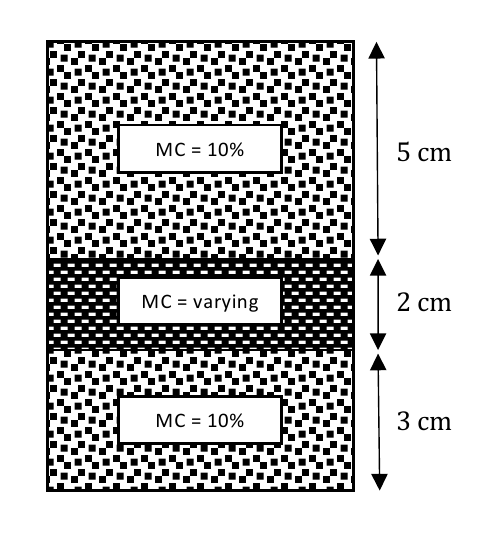}
\caption{Schematic illustration of the one-dimensional computational domain with three 
layers of varying moisture content (MC).}
\label{fig:domain_layers}
\end{figure}

Critical moisture content of ignition is the moisture content above which a fuel will not 
ignite for a given boundary condition; critical moisture content of extinction is the moisture 
content above which an established smoldering front does not propagate for given upstream, 
downstream, and boundary conditions.
In this section we examine whether the critical moisture contents change with fuel composition.
For this study we held density of the fuel at $\SI{200}{\kilogram\per\meter^3}$ and applied 
heat flux of $\SI{25}{\kilo\watt\per\meter^2}$ for the first $\SI{20}{\min}$ to ignite the sample.
We ran simulations at compositions 100\%, 75\%, 50\%, and 25\% cellulose and increased the 
moisture content in intervals of 10\%. To measure the critical moisture content of ignition, 
we set a uniform moisture content throughout the fuel sample.
To measure the critical moisture content of extinction, we set the top $\SI{5}{\cm}$ of the 
domain to have 10\% moisture content to ensure a self-sustained smoldering front, followed 
by a wet layer of $\SI{2}{\cm}$ whose moisture content was systematically increased to 
determine the critical moisture content of extinction, with the remaining $\SI{3}{\cm}$ 
of the sample at 10\% moisture content, as shown in Figure~\ref{fig:domain_layers}.

\begin{table}[htbp]
\centering
\tbl{Critical moisture content (MC$_c$) of ignition and extinction for different fuel compositions.}
{\begin{tabular}{@{}l c c@{}}
\toprule
\% Cellulose & MC$_c$ of ignition & MC$_c$ of extinction  \\ 
\midrule
25 & 40 & 70 \\
50      & 30 &  60 \\
75 & 30 & 60 \\
100      & 30 &  60 \\
\bottomrule
\end{tabular}}
\label{crtical_moisture_content_ext}
\end{table}

Table~\ref{crtical_moisture_content_ext} shows how fuel composition affects the critical 
moisture contents of ignition and extinction.
For all compositions, critical moisture content of ignition is always lower than critical 
moisture content of extinction. Neither critical moisture content is sensitive to fuel 
composition until the mixture contains 75\% hemicellulose, when both critical moisture 
content of ignition and extinction increase by 10\%.
As previously shown in Fig.~\ref{temp_comp_dens}, adding hemicellulose to the fuel 
increases the mean peak temperature.
At this composition, the fuel samples become hot enough to sustain smoldering 
combustion even at 10\% higher moisture content.

\begin{figure}[htbp]
\centering
\includegraphics[width= 0.9\linewidth]{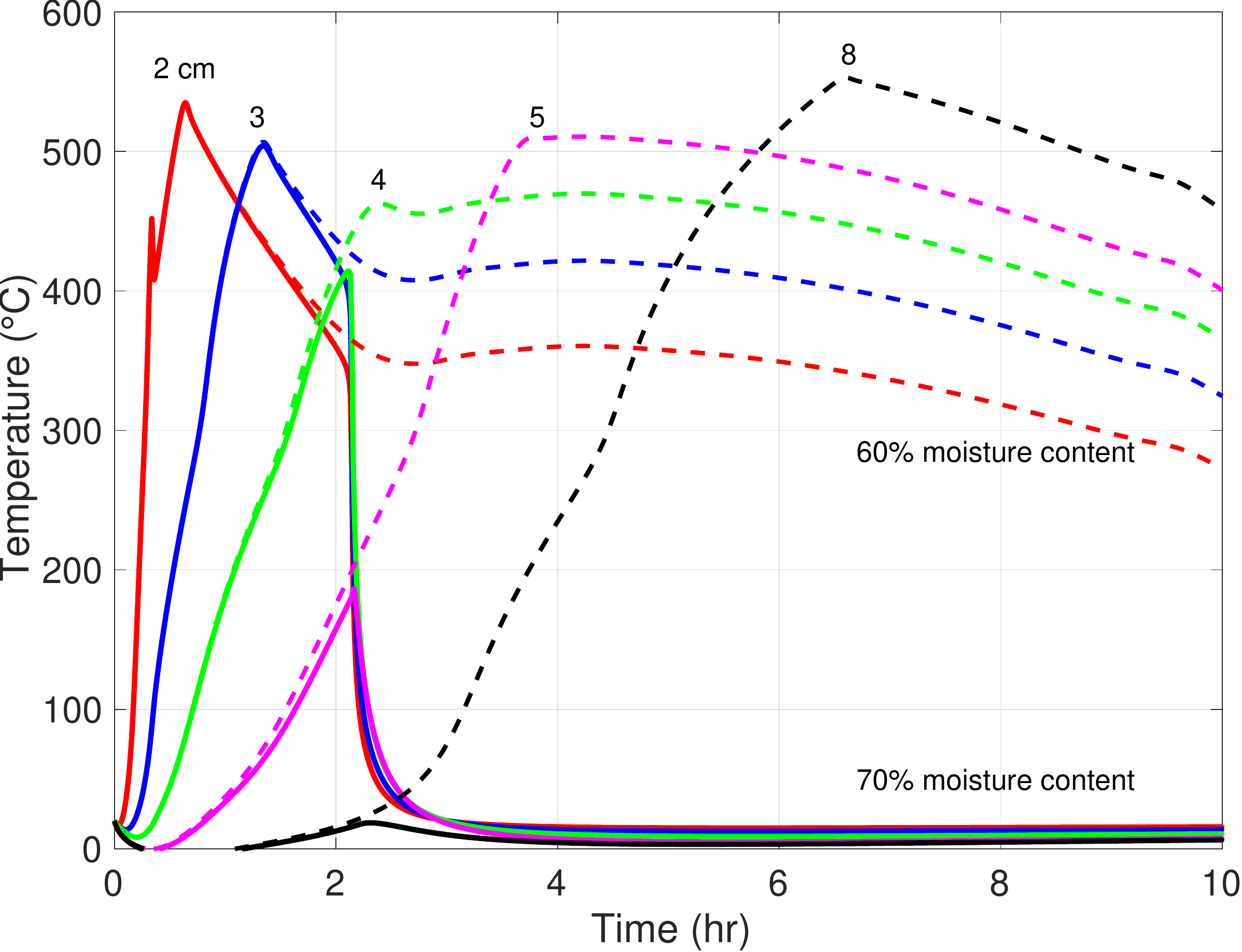}
\caption{Temperature profiles of 100\% cellulose with moisture content of wet layer 60\% 
shown by dashed line and 70\% shown by solid line at various depths.}
\label{temp_prof_ext}
\end{figure}

Figure~\ref{temp_prof_ext} shows the temperature profiles at different depths for fuel samples when the moisture content of the wet layer is 60\% and 70\%; for 60\%, smoldering propagates through the wet layer, but at 70\% smoldering combustion extinguishes.
At \SI{2}{\centi\meter} deep, the peak temperatures of the two cases match.
However, as the smoldering fronts progress deeper, the difference in the moisture content downstream starts affecting the temperatures from \SI{3}{\centi\meter} onward.
At \SI{4}{\centi\meter} deep the temperature of the 70\% moisture content case drops below the point where smoldering cannot self-sustain and it extinguishes.
On the other hand, the sample with 60\% moisture content has a peak temperature just below \SI{500}{\degreeCelsius} at \SI{4}{\centi\meter}, which is high enough to sustain smoldering.
The biggest drop in the peak temperature, for the case where there was self-sustained smoldering, is approximately \SI{1}{\centi\meter} above the point where the wet layer begins and not at the point of wet layer. 
This is because, as observed in Fig.~\ref{rr_cell_hemi}, the drying process starts before char oxidation reactions using the heat liberated from char oxidation reaction along the depth of the fuel. 
So in this particular case, the drying of the wet layer began when char oxidation reactions were occurring approximately~\SI{1}{\centi\meter} above the wet layer.

\section{Conclusions}
\label{p2:Conclusion}

In this work, we updated a one-dimensional computational model for smoldering
combustion of cellulose and hemicellulose mixtures using the open-source
software Gpyro. The model successfully predicts results from experiments 
at four fuel densities and compositions.
We used the model to examine the impact of changing fuel composition and
density on smoldering propagation speed and peak temperature. 
We also examined the role of moisture content, and how fuel composition
affects critical moisture content of ignition and extinction.

As the density of the fuel increases, the mean propagation speed drops.
This is caused by the increase in the amount of fuel that needs to be
converted to ash, which slows fuel shrinkage and thus access to oxygen.
In contrast, propagation speed increases with hemicellulose content in 
the fuel, due to the faster pyrolysis of hemicellulose compared with
cellulose. Mean peak temperature also increases with additional 
hemicellulose content, caused by the formation of ash with lower 
radiation loss across pores. Mean peak temperature increases with 
increasing density, due to decreasing radiation losses across the 
pores of the fuel.

When moisture content is added and the fuel is allowed to expand, 
the propagation speed increases due to the reduction in density.
If the fuel does not expand with the addition of water (i.e., moisture 
simply fills the pores), propagation speed drops primarily due to the
increase in wet bulk density. 
Therefore, accurately modeling smoldering in a given fuel requires 
characterizing whether moisture content causes expansion.
In both cases, additional moisture content reduces the mean peak temperature slightly.
Fuel composition increases the critical moisture content of ignition and extinction 
only when hemicellulose becomes the major constituent, due to larger heat release.

Future studies should focus on generalizing the model 
to consider lignin, the third important component of biomass and woody fuels.
In addition to validating a general fuel model with global outputs such as 
propagation speed and peak temperature, model outputs should be compared with experimental measurements 
of temperature profiles and mass to further-constrain the model.
In addition, the impact of material and kinetic parameter
uncertainty on quantities of interest should be studied.
Based on the range of thermophysical parameter values (i.e., $\rho_{s,i}$, $k_{s,i}$, $c_i$)
found in the literature~\cite{Miller_Bellan_1997,Mwaikambo_Ansell,Qi2020,Richter2021,Yan2019,Huang2016,Huang2017},
initial estimations suggest an uncertainty of 5--8\% in propagation speed and 3\% in peak
temperature, which warrants a more-complete uncertainty analysis.

\section*{Acknowledgements}
We also thank David Blunck, Benjamin Smucker, and Daniel Cowan at Oregon State University for providing their experimental temperature measurements data for validation.

\section*{Funding}
This research was funded by the Strategic Environmental Research and Development Program (SERDP) award RC-2651 under contract number W912HQ-16-C-0045.

The views, opinions, and/or findings contained in this report are those of the authors and should not be construed as an official Department of Defense position of decision unless so designated by other official documentation.

\section*{Supplementary material}
The supplementary material for this article contains the complete kinetic
parameters used in the model, along with additional details about the experiments used to validate the model.
In addition, all of the Gpyro input files, plotting scripts, and figures 
are available openly under the CC-BY license~\citep{repropack}.
Gpyro itself is available openly \citep{gpyro:0.7}.


\bibliographystyle{tfq}
\bibliography{references.bib}

\end{document}